\documentclass[utf8]{FrontiersinHarvard} 

\usepackage{url,hyperref,lineno,microtype,subcaption}
\usepackage[onehalfspacing]{setspace}
\usepackage{float}

\newcommand{\bl}[1]{\mbox{\boldmath$ #1 $}}

\def\keyFont{\fontsize{8}{11}\helveticabold }
\def\firstAuthorLast{Vorobyov {et~al.}} 
\def\Authors{Eduard I. Vorobyov\,$^{1,2,3,*}$, Aleksandr M. Skliarevskii\,$^{3}$, Vardan G. Elbakyan\,$^{4,3}$, Michael Dunham\,$^{4}$ and Manuel G\"udel\,$^{1}$}
\def\Address{$^{1}$University of Vienna, Department of Astrophysics, T\"urkenschanzstrasse 17, 1180, Vienna, Austria; \\
$^{2}$Institut für Astro- und Teilchenphysik, Universität Innsbruck, Technikerstraße 25, 6020 Innsbruck, Austria \\
$^{3}$Research Institute of Physics, Southern Federal University, Rostov-on-Don 344090, Russia; \\
$^{4}$Fakultät für Physik, Universität Duisburg-Essen, Lotharstraße 1, D-47057 Duisburg, Germany \\
$^{5}$Department of Physics, Middlebury College, Middlebury, VT 05753, USA }

\def\corrAuthor{Eduard Vorobyov}

\def\corrEmail{eduard.vorobiev@univie.ac.at}

\begin{document}
\onecolumn
\firstpage{1}

\title[Disk masses and sizes]{On the accuracy of mass and size measurements  of young protoplanetary disks} 

\author[\firstAuthorLast ]{\Authors} 
\address{} 
\correspondance{} 

\extraAuth{}

\maketitle

\begin{abstract}

Knowing the masses and sizes of protoplanetary disks is of fundamental importance for the contemporary theories 
of planet formation. However, their measurements are associated with large uncertainties. In this proof of concept study, we focus on the very early stages of disk evolution, concurrent with the formation of the protostellar seed, 
because it is then that the initial conditions for subsequent planet formation are likely established.
Using three-dimensional hydrodynamic simulations of a protoplanetary disk followed by radiation 
transfer postprocessing, we constructed synthetic disk images at millimeter wavelengths. We then calculated the synthetic disk radii and masses using an algorithm that is often applied to observations of protoplanetary disks with ALMA, and compared the resulting values with the actual disk mass and size derived directly from hydrodynamic modeling.  We paid specific attention to the effects of dust growth on the discrepancy between synthetic and intrinsic disk masses and radii.
We find that the dust mass is likely underestimated in Band~6 by factors of 1.4--4.2 when Ossenkopf \& Henning opacities and typical dust temperatures are used, but the discrepancy reduces in Band~3, where the dust mass can be even overestimated. Dust growth affects both disk mass and size estimates via the dust-size-dependent opacity, and extremely low values of dust temperature ($\approx$ several Kelvin) are required to recover the intrinsic dust mass when dust has grown to mm-sized grains and its opacity has increased. Dust mass estimates are weakly sensitive to the distance to the source, while disk radii may be seriously affected.  
We conclude that the accuracy of measuring the dust mass and disk radius during the formation of a protoplanetary disk also depends on the progress in dust growth. The same disk, but observed at different stages of dust growth and with different linear resolutions, can have apparent radii that differ from the intrinsic value by up to a factor of two. Multi-wavelength observations that can help to constrain the maximum dust size would be useful when inferring the disk masses and sizes.

\tiny
 \keyFont{ \section{Keywords:} Protostellar disks, Protoplanetary disks, Hydrodynamics, Stars: formation, dust evolution} 
\end{abstract}

\section{Introduction}

Knowing the masses of protoplanetary disks is of fundamental importance for the contemporary theories of planet formation. The mass of gas in the disk influences its tendency to undergo gravitational instability and form planets through disk gravitational fragmentation \citep{2007MayerLufkin,2013Vorobyov,2017Nayakshin,2020Stamatellos,2023Boss,2025ApJ...986...91X}. The mass of dust in the disk sets the upper limits on the masses of terrestrial planets and cores of giant planets to be formed via planetesimal hierarchical growth or pebble accretion \citep{Pollack1996,Lambrechts2012,Jin2014}. Dust particles at the upper part of the size spectrum are the main carriers of dust mass in the disk \citep{2016Birnstiel}, but they also contribute strongly to the (sub-)millimeter opacity in the outer cold and gravitationally unstable disk regions if their size exceeds a certain threshold value \citep{Pavlyuchenkov2019}.

Knowing the sizes of protoplanetary disks is of no less importance for planet formation theories. 
Gravitational fragmentation is known to operate in the disk at distances beyond tens of astronomical units, because at smaller distances the high rate of shear and slow cooling prevents disk fragmentation \citep{Gammie2001,Rice2003,Meru2012}. The size of the dust disk, which is usually smaller than that of the gas disk\citep{Ansdell2018,Trapman2019,Hsieh2024}, carries information about the efficiency of radial dust drift and, indirectly, about the efficiency of dust growth in the system.  The size of the dust disk also defines the spatial extent within which planetesimals -- the first building blocks of protoplanets -- are expected to form.  Interestingly, recent observational data on FU Orionis-type objects  revealed several features that do not fit into the contemporary models of disk evolution, dust growth, and planet formation, indicating that the disks of these outburst objects tend to be of a smaller size but higher mass than disks around their quiescent counterparts \citep{Kospal2021}.  The time evolution of both masses and sizes reflects the dominant transport and loss mechanisms of mass and angular momentum in a protoplanetary disk \citep{Manara2023}.

While the masses and sizes of the gas disk  are usually obtained via the observations of CO isotopologues, sometimes complemented with N2H$^+$, or hydrogen deuteride (HD) \citep{McClure2016,Trapman2025}, the dust disk is probed via observations of thermal dust emission at (sub-)mm wavelengths \citep{Tobin2020,Tychoniec2020,2022ApJ...938...55A}. 
Observational estimates of gas masses suffer from uncertainties in the abundance of gaseous tracers, their freezeout on dust grains, and strong dependence of population levels on the gas temperature \citep{Bergin2017,2017ApJ...849..130M}. Alternatively, dynamical measurements that involve fitting of the disk rotation curve to the Keplerian law can provide estimates on the gas mass in the disk \citep{Veronesi2021}, but such exercises require high resolution observations.

Estimates of dust mass in the disk are no less affected by uncertainties in the dust opacity, inclination, and dust temperature  \citep{2014DunhamVorobyov,Tobin2020,Tychoniec2020}, particularly, if the optically thin assumption is made.  The latter may be justified for evolved Class II systems, but not for younger Class 0/I counterparts.  The situation may become even more complicated in protostellar systems that are in their earliest stages of formation. Observationally, these objects may reveal themselves as very low luminosity objects or VELLOs \citep{2010Dunham, 2019ApJS..240...18K}, which may be the first hydrostatic cores (FHSCs) surrounded by nascent disks \citep{Vorobyov-VELLO-2017}. Indeed, numerical hydrodynamics simulations indicate that a protoplanetary disk may start forming before the FHSC collapses due to molecular hydrogen dissociation to form the protostellar seed \citep{Inutsuka2012, Tomida2015,VorobyovKulikov2024}.

In this work, we aim to determine the accuracy of the dust mass and disk radius measurements in a protoplanetary disk in its very early stages that are concurrent with the formation of the protostellar seed. This is important because this stage is likely to set the initial dust repository for subsequent planet formation.
To this end, we produce synthetic observables of the protoplanetary disk made with a dust distribution derived from three-dimensional numerical hydrodynamics simulations of cloud collapse and disk formation.
We also consider models with artificially imposed limits on the maximum dust size to account for electrostatic or bouncing barriers, which are not considered self-consistently in the dust growth model \citep{Vorobyov2025}.
This allows us to connect numerical simulations with synthetic observations and test the basic assumptions underlying dust mass and disk radius estimates in real observations. 
In Sect.~\ref{Sect:method} the numerical hydrodynamics model is briefly reviewed. The methods to estimate disk masses and sizes from numerical data are described in Sect.~\ref{Sect:mass-size}. The main results are presented in Sect.~\ref{Sect:results}. Model limitations and comparison with previous work are provided in Sect.~\ref{Sect:discuss}. Main conclusions are summarized in Sect.~\ref{Sect:conclude}.

\section{Numerical model}
\label{Sect:method}
In this section, we provide a basic description of the numerical hydrodynamics model employed to simulate the formation of a protoplanetary disk. We used the three-dimensional ngFEOSAD code to simulate the gravitational collapse of a pre-stellar cloud and early evolution of a nascent protoplanetary disk. The code solves the equations of gas and dust dynamics including self-gravity and dust growth in the polytropic approximation on nested Cartesian meshes. The detailed description of the code and the pertinent equations can be found in \citet{VorobyovKulikov2024}, and also in a concise form in Appendix~\ref{App:equations}. Here, we provide the basic information that is relevant to calculating the synthetic observables of our model protoplanetary disk.

The numerical simulations start from the gravitational collapse of a slowly rotating Bonnor-Ebert sphere with a mass of $0.87~M_\odot$. The initial temperature of the cloud is 10~K and it is given an initial positive density perturbation of 30\% to initiate the collapse. The ratios of rotational-to-gravitational and thermal-to-gravitational energies are 0.5\% and 70\%, respectively. Initially, the dust-to-gas mass ratio is set equal to 1:100 and the maximum dust size is $a_{\rm max}=2.0$~$\mu$m. The minimum dust size is kept fixed at $5.0\times10^{-3}$~$\mu$m and the slope of the dust size distribution is kept constant at $p=-3.5$ throughout the simulation for simplicity.

The dust-to-gas mass ratio throughout the computational domain begins to deviate from the initial value as collapse proceeds, the disk forms, and dust starts growing and settling to the disk midplane. 
Dust enhancement of the innermost cloud regions occur already in the predisk stage due to differential collapse of gas and dust in the cloud \citep{Bate2022}. Mild dust growth also occurs in the collapse stage \citep{Vorobyov2025}, but its main phase begins when the FHSC forms and the disk begins to build around the FHSC owing to conservation of angular momentum of infalling matter.  We terminate the simulations just before the FHSC is about to collapse due to molecular hydrogen dissociation and to form the protostellar seed. We therefore address the very early stages of evolution when the luminosity of the central source is still low but the protoplanetary disk may already have started to form.
Considering these early stages of disk evolution also makes it easier to model the synthetic disk images because uncertainties with the luminosity and radius of a young stellar object are lifted \citep{Vorobyov-stars-2017}. We note that while compact disks around FHSCs were theoretically and numerically predicted \citep{Tomida2015,Wurster2021,Vorobyov2025}, they are still observationally elusive. The more advanced stage when the central protostar reaches a mass of $0.1-0.2~M_\odot$ will be addressed in a follow-up study.

Relevant to our model are the choices of the turbulent viscosity $\alpha_{\rm visc}=10^{-3}$ and the dust collisional fragmentation velocity $v_{\rm frag}=5.0$~m~s$^{-1}$. The former influences the dust growth efficiency and maximum dust size via the turbulent relative velocity of dust-to-dust collisions \citep{Ormel2007}, while the latter sets the fragmentation barrier and also the maximum dust size in our simplified monodisperse dust growth model \citep{Birnstiel2012}. We note that the drift barrier is self-consistently treated via the numerical solution of the dust dynamics equations with dust-to-gas friction, including the backreaction of dust on gas.  The choices of $\alpha_{\rm visc}$  and $v_{\rm frag}$ were motivated by observations of dust settling in protoplanetary disks \citep{Rosotti2023} and laboratory experiments on dust survival during collisions \citep{2015ApJ...798...34G,Blum2018}. Apart from the actual dust size distribution obtained in our numerical simulations, we also consider several models in which the maximum dust size is artificially reset to smaller values to imitate the possible effects of electrostatic and bouncing barriers \citep{2009Okuzumi,Wada2011}, not considered self-consistently in ngFEOSAD \citep{Vorobyov2025}.

We used 12 nested grids with the linear size of the outermost grid equal to 0.09~pc. The number of grid cells per Cartesian coordinate direction of each nested grid is $N = 64$. The effective numerical resolution in the inner 4.5~au is 0.14~au and it remains at a sub-au level up to 36~au. The disk at the end of simulations extends to about 50-60~au.

\section{Estimates of disk masses and sizes}
\label{Sect:mass-size}
In this section, we describe the method we used to derive the disk masses and sizes directly from our numerical hydrodynamics simulations. We also explain how we obtained synthetic observables of the protoplanetary disk with a radiation transfer tool and used these observables to compare the synthetic disk masses and sizes with those derived directly from hydrodynamic modeling.

\subsection{Deriving the disk mass and size from hydrodynamic simulations}
\label{Sect:disk-track}
The first step in our procedure is to derive the disk mass using the three-dimensional distribution of gas densities and velocities in our computational domain.
To do that in a numerical model that self-consistently computes the cloud-to-disk transition, we have to develop a means of distinguishing the disk from the infalling cloud. For this purpose, we adopted the disk tracking conditions outlined in \citet{2012Joos}. In particular, we used the following criteria to determine if a particular computational cell in the entire computational domain belongs to the disk, and not to the infalling cloud:
 \begin{itemize}
     \item the gas rotational velocity must be faster than twice the radial infall velocity, $v_{\rm \phi}>2v_{\rm r}$,
     \item the gas rotational velocity must be faster than twice the vertical infall velocity, $v_{\phi}>2v_{\rm z}$,
     \item gas must not be thermally supported, $\frac{\rho_{\rm g} v_{\phi}^2}{2}>2P$, 
     \item the gas number density must be higher than $10^{9}$~cm$^{-3}$.
 \end{itemize}
Here, $v_{\rm z}$, $v_{\rm r}$, and $v_\phi$ are the components of gas velocity in the cylindrical coordinates, $\rho_{\rm g}$ the volume density of gas, and $P$ the gas pressure.
If any of the first three conditions fails, a particular grid cell is not qualified as belonging to the disk. The forth condition must always be fulfilled. We also tried a higher threshold for the gas number density $10^{10}$~cm$^{-3}$ but found little difference.

At the earliest stages of disk evolution considered in this work ($t\le 15$~kyr after disk formation), some radial drift and vertical settling of dust towards the disk midplane may have already occurred \citep{Bate2022,Lebreuilly2020,VorobyovKulikov2024}. 
Therefore, the gas and dust disks may not be having identical extents. Since we use the volume density and velocity of gas to identify the disk-to-envelope interface, our algorithm identifies the gas disk rather than the dust disk. We neglect this difference in the present work, as we do not expect it to be decisive at this early evolution stage. Indeed, \citet{Hsieh2024} found that the ratio of gas to dust disk radii for Class~0 objects does not deviate much from unity.

\subsection{Construction of synthetic disk images in dust continuum emission}
The second step is to obtain the simulated images of our model disks in millimeter wavebands. For this purpose, we employed the RADMC-3D radiative transfer tool \citep{Dullemond2012} and postprocessed the resulting synthetic fluxes with the ALMA Observational Support Tool (ALMA OST) to account for atmospheric effects and finite interferometer resolution. 
The input parameters into RADMC-3D are the three-dimensional dust volume density, temperature, and dust size distributions obtained from ngFEOSAD and the output is the radiation intensity distribution at a specific wavelength. Since we modeled the earliest stages of evolution, the radiation of the central star was neglected.

In the RADMC-3D radiation transfer simulations we used the Cartesian grid with the mesh refinement option. It allows us to directly map the ngFEOSAD nested meshes onto the RADMC-3D grid layout.
We considered eight inner nested meshes of ngFEOSAD, which encompass a cube with an edge size of $1100$~au centered at the FHSC. Importing the dust content of the disk into RADMC-3D calculations requires specifying its density, temperature, and dust opacity. While the first two were taken directly from hydrodynamic modeling, 
the absorption and scattering dust opacities were obtained using the opTool \citep{Woitke2016} for dust grains with a fixed minimum dust size of $5\times 10^{-3}$~$\mu$m and a maximum dust size $a_{\rm max}$ (see also Appendix~\ref{App:opacity}), which value was either fixed or directly taken from our hydrodynamic simulations.
In the latter case, $a_{\rm max}$ is a continuous distribution and varies from computational cell to cell, which is too complex to implement in RADMC-3D. To circumvent this problem, we introduce in each cell a number of dust sub-populations, each characterized by the same minimum size $5\times10^{-3}$~$\mu$m but different maximum sizes: $a_{\rm max}^{\rm sub}$=2.0~$\mu$m, 10~$\mu$m, 100~$\mu$m, 1~mm, and 1~cm. We then retain in each cell only the sub-population whose maximum size $a_{\rm max}^{\rm sub}$ is closest to $a_{\rm max}$.

In each Monte-Carlo RADMC-3D simulation we set the number of photons to $N=10^8$. The synthetic radiation fluxes obtained after radiation transfer simulations are then post-processed with the ALMA OST with parameters corresponding to ALMA Band~6 or Band~3. In particular, the bandwidth is set to 7.5~GHz, the beam size is varied from $0.042^{\prime\prime} \times 0.046^{\prime\prime}$ to $0.134^{\prime\prime} \times 0.146^{\prime\prime}$, and the on-source time is equal to 1.0 hour of observation. We consider the atmospheric conditions corresponding to precipitable water vapor of 1.796~mm, which results in a theoretical RMS-noise of $\sigma \simeq 1.11\times10^{-5}$~Jy. 
We note that the adopted angular resolution is close to the highest 
limit\footnote{\href{https://almascience.nrao.edu/about-alma/alma-basics}{https://almascience.nrao.edu/about-alma/alma-basics}}, and may not be available at every ALMA cycle. 

\subsection{Estimates of disk masses and sizes from radiation fluxes}
\label{sect:mock-mass-size}

The final step is to derive the disk masses and sizes  from the radiation fluxes obtained with RADMC-3D and postprocessed with ALMA OST as described above. For this purpose, we follow an approach usually employed when inferring the masses and sizes of dust disks from observations in (sub)-mm wavebands \citep[e.g.,][]{Tobin2020,Kospal2021}.
First, we determine the radial extent within which 90\% or 97\% of the total flux at a given wavelength is confined, 
$R^{\rm obs}_{\rm 90\%}$ or $R^{\rm obs}_{\rm 97\%}$. The former value is often used in observational astronomy because the low signal-to-noise ratio near the disk outer boundary may interfere with the measurements, while the latter value may be considered as an idealized upper limit that can be achieved with high signal-to-noise observations.
The value of $R^{\rm obs}_{\rm 90\%}$  is further regarded as the nominal radius of our synthetic disk, while $R^{\rm obs}_{\rm 97\%}$ is used for comparison. 

Once the radius of our synthetic disk is calculated, the dust mass of the disk is estimated  using the following equation 
\begin{equation}
\label{Eq:mass}
    M_{\rm 90\%}^{\rm obs} = \dfrac{
         d^{2} \, F^{\rm 90\%}_{\lambda}
    }
    {
        B_{\lambda}\left( T_{\rm d} \right) \, \kappa_{\lambda}
    },
\end{equation}
where $d$ is the assumed distance to the object, $F_{\lambda}^{\rm 90\%}$ is the total flux at wavelength $\lambda$ contained within the disk extent defined by $R^{\rm obs}_{\rm 90\%}$, $B_{\lambda}\left(T_{\rm d} \right)$ is the Planck function at an average dust temperature $T_{\rm d}$, and $\kappa_{\lambda}$ is the dust  opacity per gram of gas. We also considered dust masses $M_{\rm 97\%}^{\rm obs}$ contained within 97\% of the total flux as an idealized limit. For the wavelength of observations we choose 1.3~mm, which corresponds to the center of the B6~band on ALMA. The dust opacity $\kappa_{\lambda=1.3~\mathrm{mm}}$ is set equal to 0.89 cm$^2$~g$^{-1}$  for $\lambda=1.3$~mm \citep{OssenkopfHenning1994}, which is a frequent observer's choice \citep{Tobin2020,Kospal2021}, and is scaled accordingly for $\lambda=3.0$~mm. Details on the determination of the dust temperature $T_{\rm dust}$ are provided in Sect.~\ref{Sect:disk-mass}.  We note here that Equation~(\ref{Eq:mass}) is derived in the limit of negligible dust scattering. Nevertheless, we will use in this work the synthetic fluxes and dust opacities that consider the effects of dust scattering. This is done because fluxes from real disks are by default affected by dust scattering. 
However, in Appendix~\ref{Sect:scatter} we conducted several experiments and turned off dust scattering to assess its effects on the synthetic fluxes. Equation~(\ref{Eq:mass}) is valid for the optically thin limit, but we will apply it to disk configurations that are partly or fully optically thick for the lack of a better alternative. We note that this equation is often used in large surveys of star-forming regions and can lead to serious mass miscalculations, as we will demonstrate later in this work.

\section{Results}
\label{Sect:results}
In this section, the comparison between the disk masses and sizes derived directly from hydrodynamic modeling with those derived from synthetic radiation fluxes is carried out.

\subsection{Model disk characteristics and synthetic images}

\begin{figure}[H]
    \centering
    \includegraphics[width=0.8\linewidth]{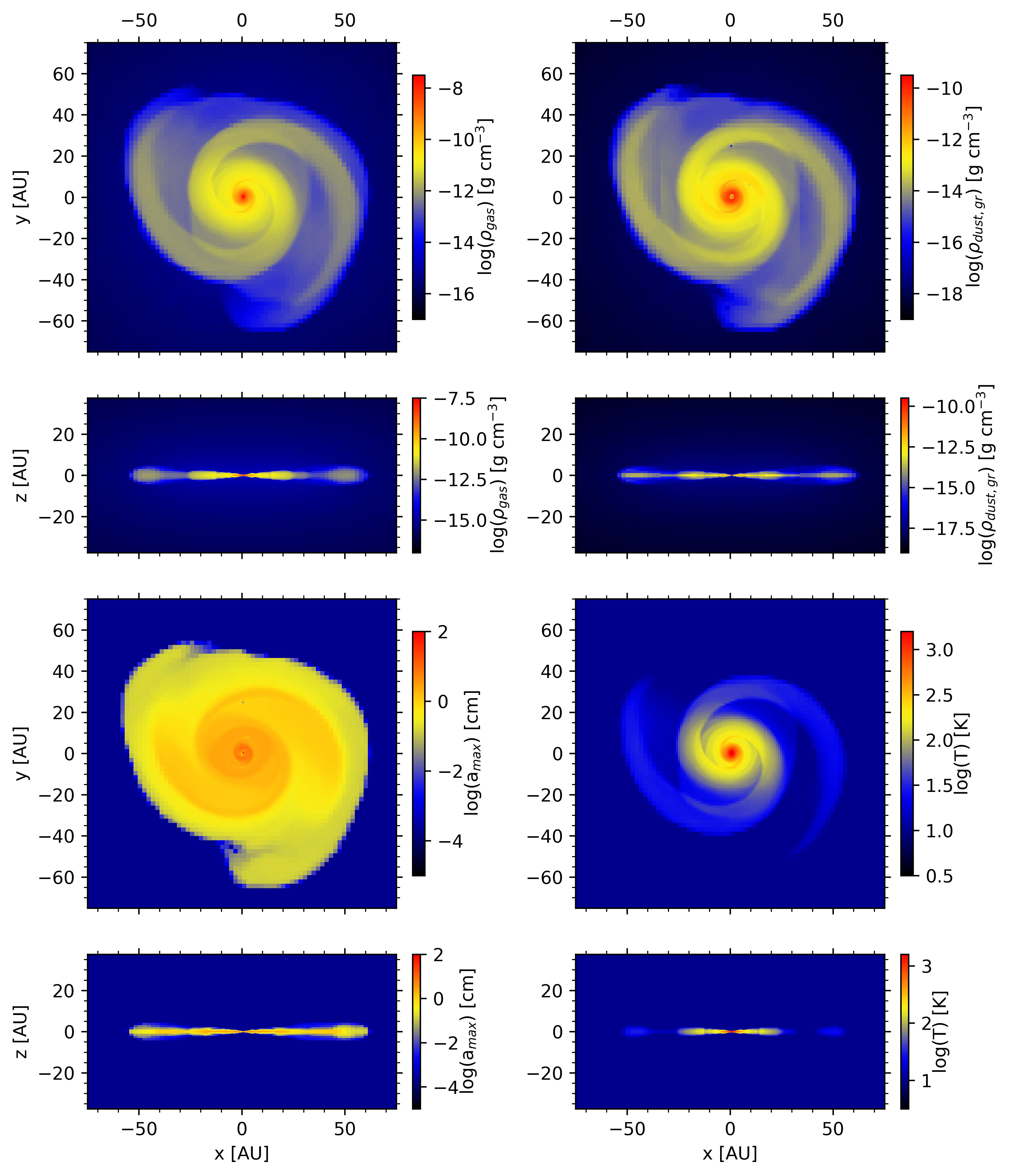}
    \caption{Model disk in the midplane and across a vertical slice. Shown are from top to bottom and from left to right: gas volume density, grown dust volume density, maximum dust size, and temperature.   }
    \label{fig:1}
\end{figure}

Figure~\ref{fig:1} presents the spatial distribution of the main disk characteristics in the disk midplane and at the vertical slice taken through the $x-z$ plane at $y=0$ as obtained from numerical hydrodynamics modeling with ngFEOSAD at 15~kyr after the instance of disk formation. A clear two-armed spiral pattern is evident in both gas and dust volume density distributions in the disk midplane.  The Toomre $Q$-parameter is less than unity in the spiral arms, indicating that the disk is strongly gravitationally unstable,  but gravitational fragmentation does not yet occur (see Appendix~\ref{App:Toomre} for more details).
The spatial distribution of grown dust in the disk midplane mostly follows that of gas. Dust evaporation in the hot center of the FHSC is accounted for (see appendix in \citet{Das2025}). The dust size in this early stage already exceeds 1.0~mm throughout most of the disk midplane, reaching a few centimeters in the innermost regions. We note, however, that fast dust growth may be somewhat tampered by electrostatic and bouncing dust growth barriers that were not considered in our numerical model \citep{Vorobyov2025}. The spiral arms are warmer than the interspiral regions and the highest temperature is achieved in the innermost regions occupied by the FHSC. 

The vertical slices reveal a radially flared gas density profile with local bulges, which correspond to the position of spiral arms. Dust settling to the disk midplane is manifested by a narrower dust density distribution than that of gas.  
The maximum dust size $a_{\rm max}$ quickly drops with increasing vertical distance from $\sim 1-10$~mm in the disk midplane down to 2.0~$\mu$m at the disk-cloud interface, a value that is characteristic of the infalling cloud in our model. 
The temperature drops with vertical distance. This may not be realistic for evolved disks that are passively heated by the central star but is justified in our case, because the radiation input from the FHSC is expected to be insignificant and the disk is mainly heated in this stage by turbulent viscosity and PdV work, both operating more efficiently near the disk midplane, rather than in the disk atmosphere.

\begin{figure}[H]
\begin{centering}
\includegraphics[width=1.0\linewidth]{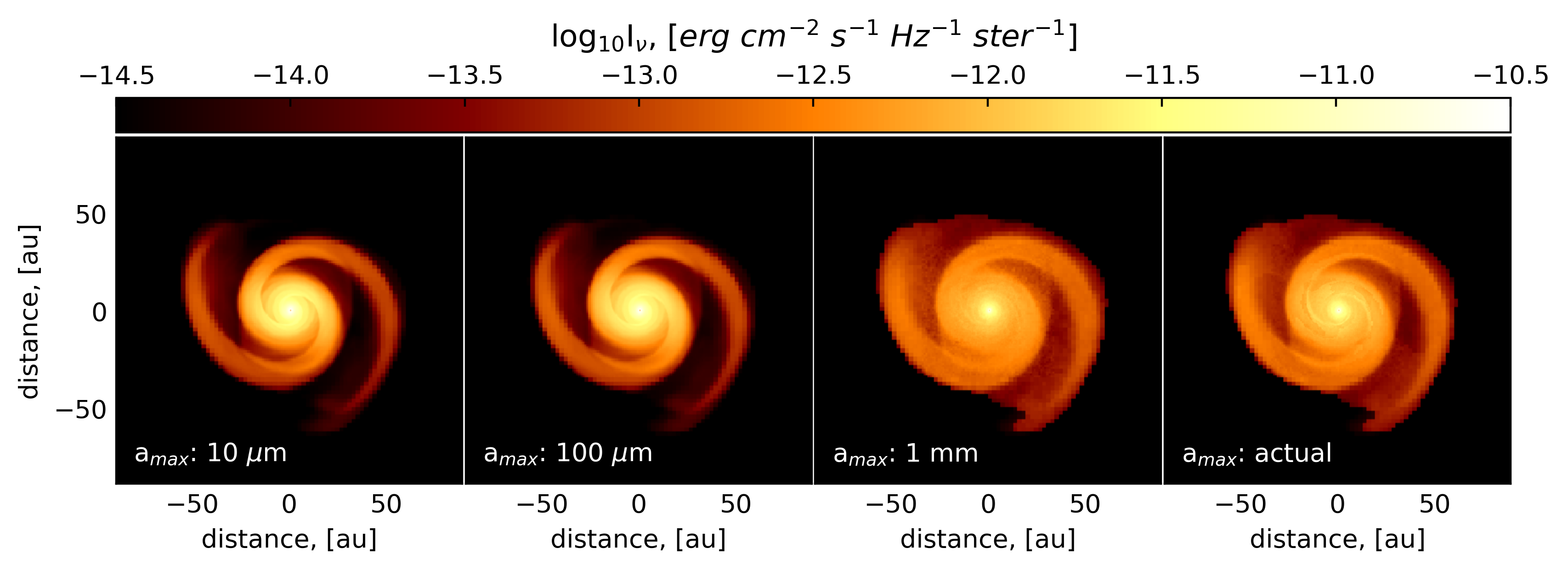} 
\par \end{centering}
\caption{Synthetic intensity distributions assuming different maximum size of the dust grains in the disk. From left to right: $a_{\rm max} = 10 \ \mu$m, $100 \ \mu$m, 1~mm, and spatially varying $a_{\rm max}$ distribution, directly taken from simulations. The envelope has  $a_{\rm max} = 2 \ \mu$m in all models considered.
} 
\label{fig:2}
\end{figure}

\begin{figure}[H]
\begin{centering}
\includegraphics[width=1.0\linewidth]{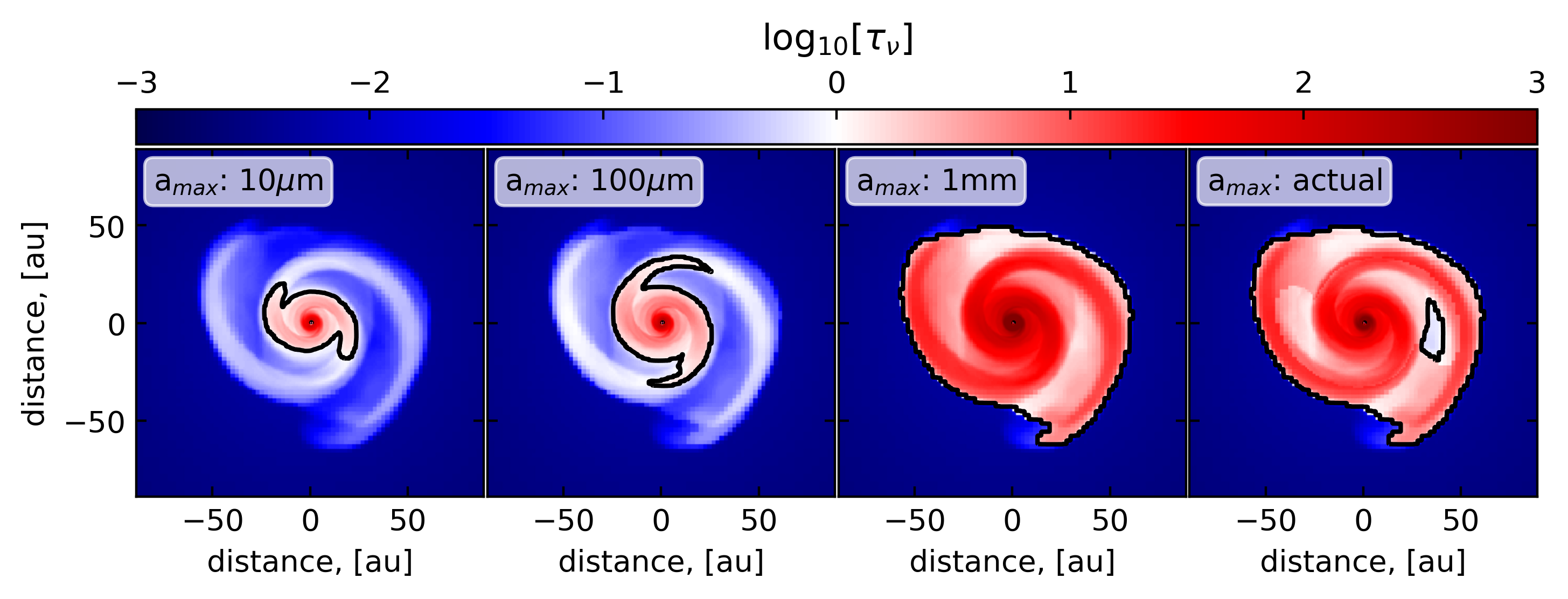} 
\par \end{centering}
\caption{Optical depth in models with different maximum dust sizes in the disk. From left to right are the cases with $a_{\rm max}=10$~$\mu$m, $a_{\rm max}=100$~$\mu$m, $a_{\rm max}=1.0$~mm, and actual model distribution. The maximum dust size in the envelope is 2.0~$\mu$m in each case.} 
\label{fig:3}
\end{figure}

Figure~\ref{fig:2} presents synthetic radiation intensity images at 1.3~mm obtained by postprocessing the model disk shown in Figure~\ref{fig:1} with RADMC-3D. The simulation box is $1100\times 1100 \times 1100$~au$^3$, comprising the inner eight nested meshes of the entire numerical hydrodynamics domain, but we show only the inner $90\times 90$~au$^2$ region with a face-on orientation of the disk. To assess the effect of dust growth on the synthetic disk images and to account for the possible effects of electrostatic and bouncing barriers, we considered several  cases of the dust size distribution across the disk. Apart from the dust size distribution directly obtained from hydrodynamic modeling with ngFEOSAD (see Figure~\ref{fig:1}) and characterized by spatially varying values of the maximum dust size $a_{\rm max}$ (hereafter, referred to as ``actual'' distribution), we also considered several fixed values for $a_{\rm max}$. In particular, we have artificially reset the actual dust sizes throughout the disk to a constant value, keeping the gas and dust densities  unchanged. We note that we do not recompute the entire model with fixed dust sizes, but rather reset the maximum dust size alone.

Figure~\ref{fig:2} shows the results of our experiments with a fixed $a_{\rm max}=2.0$~$\mu$m in the cloud, but different $a_{\rm max}$ in the disk (note that the minimum size stays constant at $a_{\rm min}=5\times10^{-3}$~$\mu$m in all cases). As the maximum size of dust in the disk increases, the appearance of the spiral arms in synthetic radiation intensity images becomes more diffuse. The transition is clearly evident between $a_{\rm max}=100$~$\mu$m and 1.0~mm. More specifically, the transition occurs between $a_{\rm max}=200~\mu$m and 300~$\mu$m, but the intensity maps at these wavelengths are similar to the images at $a_{\rm max}=100$~$\mu$m and 1.0~mm, respectively. This effect is related to a sharp increase in the dust opacity above a certain maximum dust size, defined as $a_{\rm max}^{\rm crit} = \lambda / (2 \pi)$, where $\lambda$ is the wavelength of observation (see Appendix~\ref{App:opacity}). We note that dust of different size may also have different dynamics and efficiency of dust accumulation in spiral arms, an effect that is not considered in our controlled experiments. However, as was demonstrated in \citet{VorobyovKulikov2024}, dust trapping in spiral arms is greatly reduced due to low Stokes numbers in warm and dense environments of protoplanetary disks in the earliest evolutionary stages, an effect which is indirectly confirmed by the lack of spiral patterns in deeply embedded disks \citep{Ohashi2023}.

The effect of sharp increase in opacity is corroborated in Figure~\ref{fig:3} showing the optical depth of our model disk at $\lambda=1.3$~mm including contributions from absorption and isotropic scattering.  For the case of $a_{\rm max}$=10~$\mu$m and 100~$\mu$m, the outer spiral arms and the interspiral regions are optically thin. Only the FHS, the inner disk regions and the inner parts of spiral arms are optically thick. We note, however, that these optically thick regions may contain a significant fraction of the total radiation flux (see Fig.~\ref{fig:R-disk2}). 
The optically thin medium effectively means that the radiation intensity is proportional to the product of the optical depth and the Planck function. For a larger $a_{\rm max}=1$~mm and for the actual maximum dust size distribution, the optical depth is above 1.0 almost throughout the entire disk extent, apart from isolated interarmed regions, and the radiation intensity in these optically thick cases is mostly determined by the Planck function. We note also that the spiral arms appear slightly sharper when the actual dust sized distribution is applied, rather than that with a fixed upper limit of $a_{\rm max}=1.0$~mm.

\begin{figure*}
\begin{centering}
\includegraphics[width=0.85\linewidth]{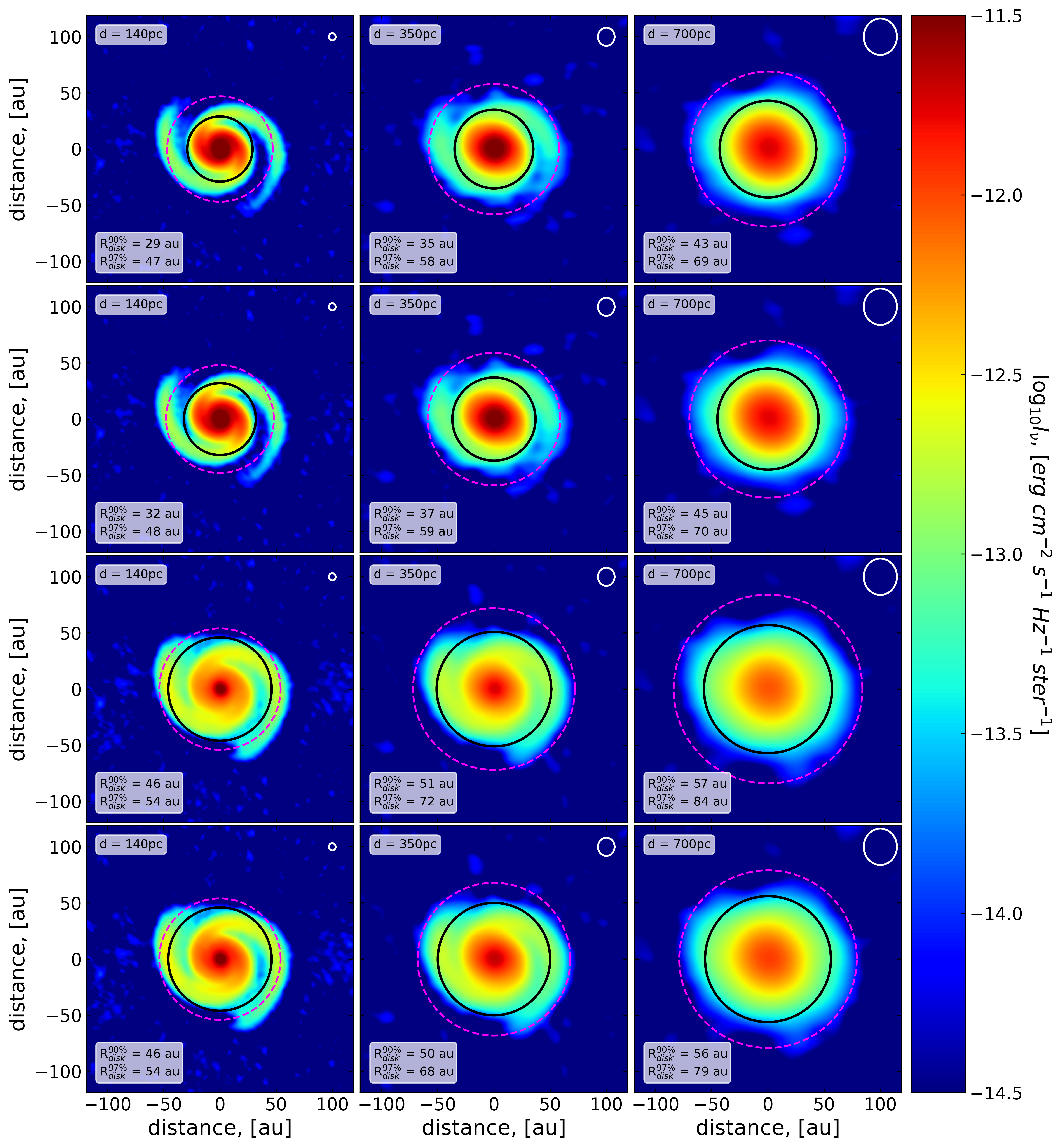}
\par \end{centering}
\caption{Synthetic intensities of the model disk at 1.3~mm obtained with RADMC-3D and postprocessed using the ALMA OST at Band 6.  The three columns show, from left to right, the three assumed distances to the object: 140~pc, 350~pc, 700~pc. The rows from top to bottom are models with: $a_{\rm max}^{\rm disk} = 10 \ \mu$m, $100 \ \mu$m, $1$~mm, and actual maximal dust size in the disk.
The black and red circles cover the regions of the disk where 90\%  and 97\% of the total radiation flux is contained, defining the disk sizes according to the adopted criteria. In all cases, $a_{\rm max}$ in the cloud is 2.0~$\mu$m.  The white circles in the top-right corner of each panel show the linear size of the beam. } 
\label{fig:6alma}
\end{figure*}

\subsection{Postprocessed synthetic disk images in Band 6}
\label{Sect:disk-size}
As a next step, we take the synthetic disk images shown in Figure~\ref{fig:2} and postprocess them with the ALMA observational support tool (OST) \citep{ALMA-OST2011} to add atmospheric noise and finite resolution effects. Figure~\ref{fig:6alma} depicts the resulting disk images in Band~6 of ALMA  with a beam size of $0.042^{\prime\prime} \times 0.046^{\prime\prime}$  and an exposure time of 1.0 hour. The chosen configuration provides an angular resolution that is a factor $\approx 2.3$ lower than the maximum achievable resolution of ALMA in Band~6\footnote{\href{https://almascience.eso.org/about-almas}{https://almascience.eso.org/about-alma}}. Three different distances to the object are chosen: $R=140$~pc, 350~pc, and 700~pc,  which correspond to the linear sizes of the beam: $\approx$~6.0~au, 15~au, and 30~au. Varying the distance to the source but keeping the angular resolution fixed at $0.042^{\prime\prime} \times 0.046^{\prime\prime}$  would imitate the observations of distinct star-forming clusters with the same configuration of ALMA antennas (e.g., Cycle 11, C-8). Each row of panels correspond to dust size distributions in the disk that are characterized by different maximum dust sizes $a_{\rm max}$ but similar $a_{\rm min}$, both chosen in accordance with Figure~\ref{fig:2}. 
In the infalling cloud, the dust size spectrum is similar in all cases, with minimum and maximum dust sizes of $5\times10^{-3}$~$\mu$m and 2.0~$\mu$m, respectively.
We also note that in some regions beyond the disk, where the detection limit was lower than noise, negative fluxes have occurred. These regions were plotted with the color corresponding the lowest detected limit, $\log I_\nu = -14.5$~in CGS units.

Several interesting features can be noted from our mock observations. Firstly, the spiral arms smear out with increasing distance and are visually indistinguishable at $R = 700$~pc. 
The spiral pattern also becomes less sharp with increasing $a_{\rm max}$, as was already noted for the unprocessed radiation intensities in Figure~\ref{fig:2}. This latter finding implies that the spiral pattern caused by gravitational instability may be easier to detect in younger disks with $a_{\rm max}\le 100~\mu$m where the main phase of dust growth is yet to occur (see also \citealt{Hall2019}). We also note that the fiducial criterion for the disk radius ($R^{\rm obs}_{\rm 90\%}$)  cuts out a large portion of the spiral arms in the disk if the main dust growth phase has not yet occurred (first and second rows in Fig.~\ref{fig:6alma}).

Secondly, the synthetic disk radii ($R^{\rm obs}_{\rm 90\%}$ and $R^{\rm obs}_{\rm 97\%}$) grow with increasing distance to the source. This may be the effect of beam smearing, as the linear size of the beam notably increases with distance, see the white circles in Figure~\ref{fig:6alma}.  The effect is strongest for $a_{\rm max}\le$~100~$\mu$m with an increase in $R^{\rm obs}_{\rm 90\%}$ and $R^{\rm obs}_{\rm 97\%}$ of up to $50\%$, but even for the mm-cm sized grains the increase can be up to 20\%.  We note that subtracting about 1/2 of the linear size of the beam from the synthetic disk radii can help to compensate for the effects of beam smearing (with increasing distance) on the apparent radius of the dust disk, in particular for $R^{\rm obs}_{\rm 90\%}$.

Thirdly, disks in a more advanced stage of dust growth look somewhat bigger.  This can be related to the optical depth effects.
For the case of $a_{\rm max}\le 100$~$\mu$m,  the central optically thick  region of the disk ($\le  20$~au) dominates the radiation intensity and the optically thin spiral arms are characterized by an order of magnitude weaker values (see Fig.~\ref{fig:6alma}). For large grains and optically thick disks, the contrast between the inner and outer disk regions is reduced.

\begin{figure*}
\begin{centering}
\includegraphics[width=\linewidth]{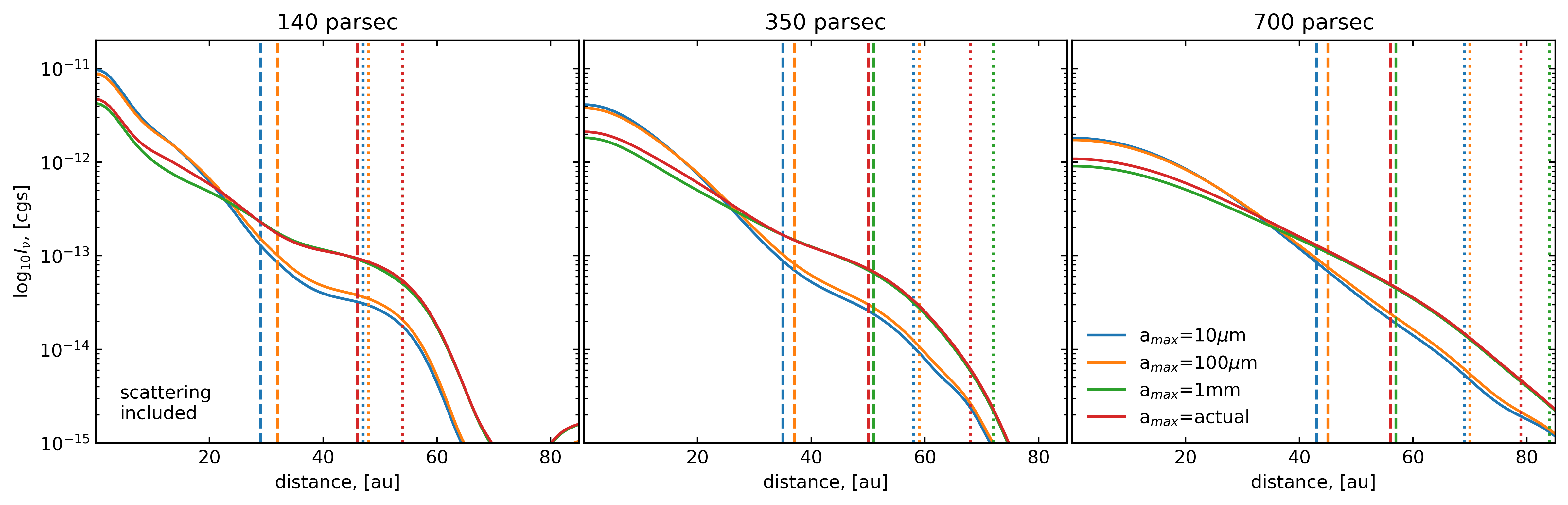} 
\par \end{centering}
\caption{Radial profiles of the radiation intensity in CGS units after postropcessing with ALMA OST. The profiles are obtained by azimuthally averaging the corresponding spatial distributions shown in Fig.~\ref{fig:6alma}. The vertical dashed and dotted lines indicate the radial positions within which 90\% or 97\% of the total flux is localizes, respectively. Panels from left to right correspond to different adopted distances to the source.} 
\label{fig:R-disk2}
\end{figure*}

Indeed, Figure~\ref{fig:R-disk2} demonstrates that the radial intensity profiles of the synthetic disk in the $a_{\rm max}\le 100$~$\mu$m limit are steep and fall off with distance rapidly.
A substantial fraction of the integrated radiation flux is, therefore, localized within the inner compact region of the disk.  
For models with $a_{\rm max}\ge1.0$~mm, however, the radiation intensity has a weaker contrast between the inner disk regions and the spiral arms, both being optically thick. As a result, the radial intensity profiles become shallower, making the disk look bigger when the same criterion (90\% or 97\%) for the integrated radiation flux is applied. At the same time, beam smearing also makes the intensity profiles look shallower as the distance to the source increases, regardless of $a_{\rm max}$.
We also note that for models with $a_{\rm max}\ge1.0$~mm the radiation intensity in the central regions of the disk decreases as compared to models with $a_{\rm max}\le 100$~$\mu$m. This is the effect of dust scattering (see Appendices~\ref{Sect:scatter} and \ref{App:opacity}), which also contributes to the overall decrease in the integrated radiation flux as discussed later in Sect.~\ref{Sect:disk-mass} in the context of dust masses.

The entire disk becomes optically thick in the $a_{\rm max}\ge1.0$~mm models   (see Figure~\ref{fig:3}) because the maximum dust size surpasses the critical value $a_{\rm max}^{\rm crit} = \lambda /(2 \pi)$. This effect is known as the opacity cliff, see also Appendix~\ref{App:opacity}. 
As dust grows and the optical depth transits from the optically thin ($\tau_\nu< 1.0$) to the optically thick ($\tau_\nu> 1.0$) regime, the radiation intensity becomes proportional to  the Planck function, rather than to the product of the Planck function and optical depth. Since both quantities decline with radius, the product of the two (optically thin limit) would produce a steeper distribution of the radiation intensity than just the Planck function (optically thick limit).
As a result, the extent within which 90\% (or 97\%) of the flux is contained (the disk radius by our definition) becomes bigger in the optically thick case.
The effect can be substantial, with an increase of more than $50\% $ in $R^{\rm obs}_{\rm 90\%}$ when the maximum dust sizes grows from $a_{\rm max}$=10~$\mu$m to mm-cm sized grains in the disk. 
The trend is also present but less expressed when $R^{\rm obs}_{\rm 97\%}$ is used.

\begin{table*}
\center{\caption{\label{table:1}Synthetic disk radii in Band 6}}
\begin{tabular}{  c | c c | c c | c c }
\hline \hline
Dust size & $R_{90\%}^{\rm obs} / R_{97\%}^{\rm obs}$ & $\triangle R_{90\%}^{\rm obs} / \triangle R_{97\%}^{\rm obs}$  & $R_{90\%}^{\rm obs} / R_{97\%}^{\rm obs}$ & $\triangle R_{90\%}^{\rm obs} / \triangle R_{97\%}^{\rm obs}$  & $R_{90\%}^{\rm obs} / R_{97\%}^{\rm obs}$  & $\triangle R_{90\%}^{\rm obs} / \triangle R_{97\%}^{\rm obs}$ \\
$[a_{\rm max}]$ & [au] & [au] & [au] & [au] & [au] & [au]  \\ [0.5ex]
\hline \\ [-2.0ex]
10~$\mu$m & $29/47$ & -24/-6  &  $35/58$ & -18/+5 &  $43/69$ & -10/+16  \\
100~$\mu$m & $32/48$ & -21/-5 &  $37/59$ & -16/+6 &  $45/70$ & -8/+31 \\
1.0~mm & $46/54$ & -7/+1 & $51/72$ & -2/+19 & $57/84$ & +4/+31   \\
actual & $46/54$ & -7/+1  & $50/68$  & -3/+15 & $56/79$ & +3/+26   \\ [1.0ex]
\hline
\end{tabular}
\center{ \textbf{Notes.}  Each pair of columns (from left to right) presents the synthetic disk radii and their deviations  from the true value of  $R^{\rm mod}_{\rm disk}=53$~au for distances $d=140$~pc, 350~pc, and 700~pc, respectively. Plus and minus signs correspond to over- and underestimates. The slash separates values obtained with different definition of the disk radius.  } 
\end{table*}

\begin{table}
\center{\caption{\label{table:2} Integral radiation fluxes in Band 6}}
\begin{tabular}{  c c c c }
\hline \hline
Dust size & $F_{90\%}^{\rm obs} / F_{97\%}^{\rm obs}$ & $F_{90\%}^{\rm obs} / F_{97\%}^{\rm obs}$  & $F_{90\%}^{\rm obs} / F_{97\%}^{\rm obs}$\\
$[a_{\rm max}]$ & [mJy] & [mJy] & [mJy]  \\ [0.5ex]
\hline \\ [-2.0ex]
10~$\mu$m & $347 / 375$ & $57 / 62$ & $14.4 / 15.5$ \\
100~$\mu$m & $350 / 375$ & $57 / 62$ & $14.4 / 15.5$ \\
1.0~mm & $265 / 286$ & $45 / 48$ & $11.2 / 12.0$  \\
actual & $302 / 324$ & $50/ 54$ & $12.6 / 13.6$ \\ [1.0ex]
\hline
\end{tabular}
\center{ \textbf{Notes.} Second, third, and fourth columns present the fluxes at distances $d=140$~pc, 350~pc, and 700~pc, respectively. The slash separates values obtained with different definition of the flux.} 
\end{table}

\subsection{Synthetic disk masses}
\label{Sect:disk-mass}

The first step in calculating the synthetic disk masses is to determine the flux $F_{\lambda}$ in Equation~(\ref{Eq:mass}). 
Table~\ref{table:2} presents the integrated synthetic fluxes contained within  $R^{\rm obs}_{\rm 90\%}$ and $R^{\rm obs}_{\rm 97\%}$ for all three adopted distances. As expected, $F_{90\%}^{\rm obs}$ and $F_{97\%}^{\rm obs}$ decrease with distance, but their behavior with increasing  $a_{\rm max}$ is more complex. The highest values are found for $a_{\rm max}\le 100$~$\mu$m and the lowest are around $a_{\rm max}=1.0$~mm. The decrease in the flux at the advanced stages of dust growth is likely caused by flux dilution due to strong dust scattering at mm-sized grains, see Appendices~\ref{Sect:scatter} and \ref{App:opacity}.

To determine dust masses from the synthetic disk images shown in Figure~\ref{fig:6alma}, we should also know the value of the average dust temperature $T_{\rm d}$ that enters Equation~(\ref{Eq:mass}). This quantity is hard to constrain directly from observations and therefore the average dust temperature is often determined for young protoplanetary disks (Class 0 and I) following \citet{Tobin2020} as:
\begin{equation}
\label{eq:t-tobin}
    T_{\rm d} = 43~K \left( {L_\ast \over 1.0 \,\mathrm{L_\odot} } \right)^{0.25},
\end{equation}
where $L_\ast$ is the luminosity of the central source. This relation is based on radiation transfer simulations of a protoplanetary disk embedded into an envelope and heated only by the central star. This case is different from the earliest evolutionary stage considered in this study, in which the disk is actively heated by internal hydrodynamic processes and stellar irradiation is neglected.   Therefore, Equation~(\ref{eq:t-tobin}) is not directly applicable in our case.

We could have used the actual hydrodynamic data to derive $T_{\rm d}$ as an alternative. However, we adopted a barotropic equation of state, which is a simplification compared to the solution of the full energy balance equation including radiation transfer.
We also note that the ngFEOSAD code makes no distinction between the gas and dust temperatures. More sophisticated simulations demonstrate that deviations may occur at the disk-envelope interface where a higher temperature of gas than that of dust may be expected due to heating by the shock wave caused by the infalling envelope \citep{Pavlyuchenkov2015,BateKeto2015,Vorobyov2020b}. 

Considering these uncertainties we decided not to focus on a particular value of the mean dust temperature but instead used a range of $T_{\rm d}=10-50$~K  when calculating the dust mass in Equation~(\ref{Eq:mass}). We note that \citet{2014DunhamVorobyov}  found that an average temperature declines from 30~K for Class 0 to 15~K for Class I objects, and an upper limit of 50~K for a very young disk considered in this work is consistent with a general temperature decline with age \citep{Commercon2012}.

\begin{table}
\center{\caption{\label{table:3}Synthetic dust masses  in Band 6 at particular $T_{\rm d}$}}
\begin{tabular}{  c c c c }
\hline \hline
Dust size & $M_{90\%}^{\rm obs}(T_{\rm d}=20$~K) & $M_{90\%}^{\rm obs}(T_{\rm d}=30$~K) & $M_{90\%}^{\rm obs}(T_{\rm d}=40$~K) \\
$[a_{\rm max}]$ & [$M_\oplus$] & [$M_\oplus$] & [$M_\oplus$]  \\ [0.5ex]
\hline \\ [-2.0ex]
10~$\mu$m & 493.5 &   297.9 & 212.8 \\
100~$\mu$m & 497.4 & 300.3 & 214.5 \\
1.0~mm & 376.2 & 227.1 & 162.2  \\
actual & 429.1 & 259 & 185 \\ [1.0ex]
\hline
\end{tabular}
\center{ \textbf{Notes.} Distance to the source is $d=140$~pc.  The intrinsic dust mass in the disk is $M_{\rm dust}^{\rm mod}=673~M_\oplus$. }
\end{table}

\begin{figure}
\begin{centering}
\includegraphics[width=\linewidth]{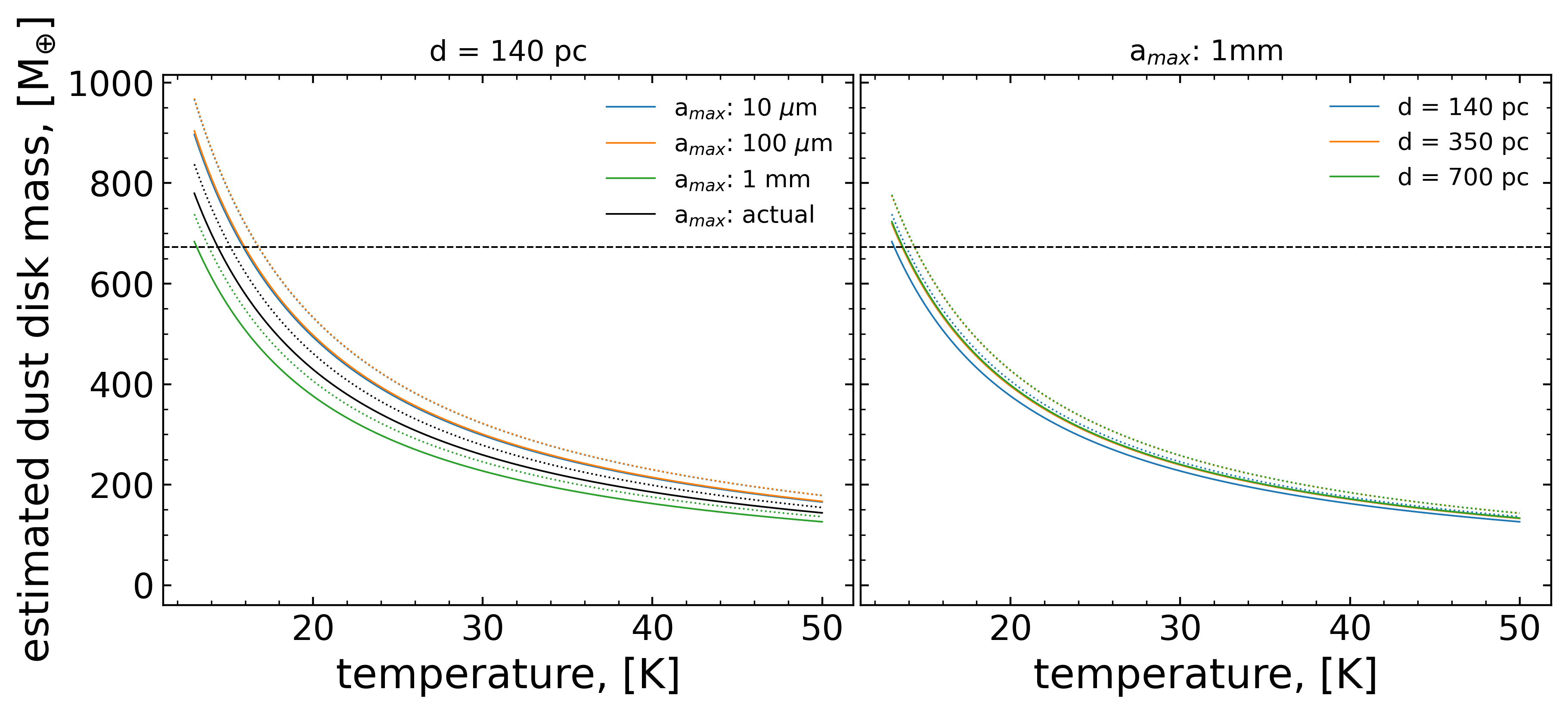} 
\par \end{centering}
\caption{Masses of dust in the synthetic disk. The left column shows the masses as a function of $a_{\rm max}$ but for the fixed distance to the source $d=140$~pc, while the right column corresponds to distinct $d$ but fixed $a_{\rm max}$=1.0~mm. Solid and dotted lines represent the mass estimates using $R^{\rm obs}_{90\%}$ and $R^{\rm obs}_{97\%}$, respectively. The horizontal dashed line defines the intrinsic dust mass in the model disk.}
\label{fig:newmass-B6}
\end{figure}

Using Equation~(\ref{Eq:mass}), we can now calculate the dust mass contained within the synthetic disk shown in Figure~\ref{fig:6alma}, with its radius defined by either $R^{\rm obs}_{\rm 90\%}$ or $R^{\rm obs}_{\rm 97\%}$.  We note that the dust opacity is that of \citet{OssenkopfHenning1994} for all calculations in this section. The resulting values are plotted in Figure~\ref{fig:newmass-B6}.
The synthetic masses of dust vary with $a_{\rm max}$ in the same manner as the integrated radiation fluxes,  namely, they decrease at the advanced stages of dust growth when $a_{\rm max}$ transits from 10~$\mu$m to 1.0~mm, likely due to flux dilution by scattering on grown dust grains. An opposite trend is observed when $a_{\rm max}$ changes from 1.0~mm to the actual dust size distribution, which is on average higher than 1.0~mm, see Fig.~\ref{fig:1}. Interestingly, the peak in the dust scattering opacity is also located around 1.0~mm, see Fig.~\ref{fig:dustOp}, thus reinforcing our conclusion about the effects of scattering on the dust mass estimates. At the same time, $M^{\rm obs}_{\rm 90\%}$ and $M^{\rm obs}_{\rm 97\%}$ only insignificantly vary with distance to the source, and this small variation is probably due to beam smearing.

We now compare the synthetic dust masses shown in Figure~\ref{fig:newmass-B6} with the actual mass of dust in our model disk. The latter is calculated by summing up all dust within the disk extent defined by the red contour line in Figure~\ref{fig:disk-R} (see Appendix~\ref{App:R-disk}). The resulting value is $M_{\rm dust}^{\rm mod} \approx 673~M_\oplus$ (evaporated dust in the FHSC is not counted), almost independent of the threshold gas number density ($10^9$ or $10^{10}$~cm$^{-3}$).   Figure~\ref{fig:newmass-B6} demonstrates a strong dependence of the synthetic dust mass on the dust temperature. In all models considered, the dust mass is heavily underestimated for $T_{\rm d}\ge20$~K by factors of several. The mismatch is somewhat sensitive to the adopted radius of the disk:  $R^{\rm obs}_{\rm 90\%}$ or $R^{\rm obs}_{\rm 97\%}$, or to the dust growth phase, but the trend to underestimate the intrinsic dust mass persists. For the reader's convenience, Table~\ref{table:3} provides the synthetic dust masses at several $T_{\rm d}$.
The underestimate of dust mass in the disk found for very young disks in our study is in agreement with similar conclusions by \citet{2014DunhamVorobyov} for Class 0 and I disks, see also Sect.~\ref{Sect:discuss}. 
A good match between the synthetic and intrinsic dust masses can only be achieved for $T_{\rm d}<20$~K, and in particular for $T_{\rm d}=$~15.9~K, 16.0~K, 13.1~K, and 14.4~K  in models with $a_{\rm max}=10$~$\mu$m, $a_{\rm max}=100$~$\mu$m, $a_{\rm max}$=1.0~mm, and in the model with the actual dust size distribution, respectively. These values were derived using $R^{\rm obs}_{90\%}$ as the disk radius. We emphasize that the best-fit temperatures are lower than what is usually assumed when recovering dust masses from observations, $T_{\rm d}=20-40$~K \citep{Ansdell2017,Tobin2020,Kospal2021}, but are comparable to the average temperature in the envelope and the optically thin disk regions.
As we will see below, the proper choice of average dust temperature may also depend on $a_{\rm max}$ and the ALMA band.

\begin{figure}  
\begin{centering}
\includegraphics[width=0.8\linewidth]{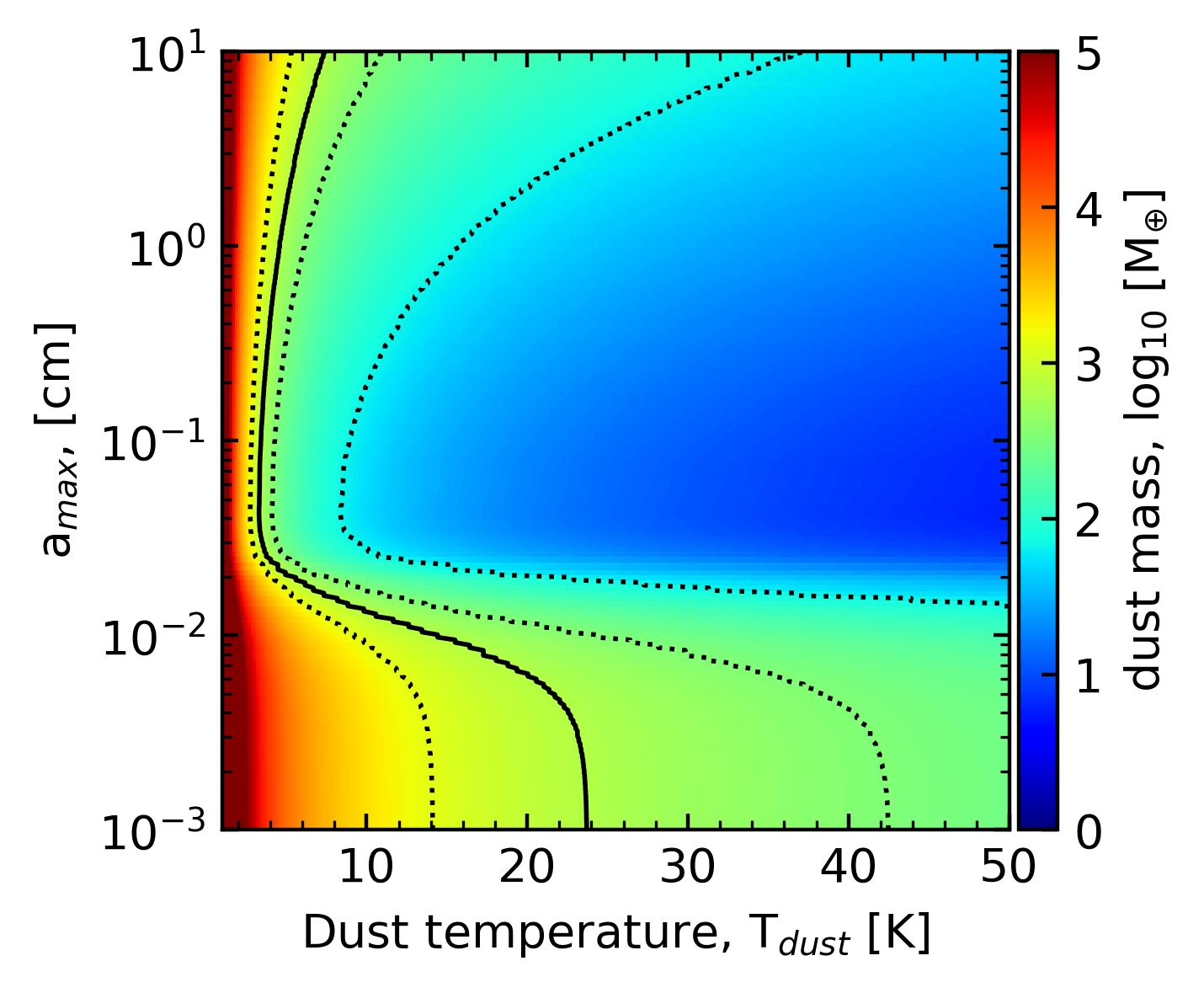}
\par \end{centering}
\caption{ Synthetic dust mass in the disk, $M_{\rm 90\%}^{\rm obs}$, as a function of dust temperature and maximum dust size. Radiation fluxes for the actual dust size distribution in Band~6 were used to derive the masses. 
The solid line indicates the true dust mass in the disk, $M_{\rm disk}^{\rm mod}=673~M_\oplus$.  The dotted lines delineate the synthetic dust masses that differ by factors 2.0, 0.5, 0.1 from the true disk mass.
} 
\label{fig:dust-mass}
\end{figure}

\subsection{The effect of dust-size-dependent opacities }

The value of the mean dust temperature depends on the intricacies of the averaging procedure \citep{Tobin2020} and on the disk evolutionary stage \citep{2014DunhamVorobyov,Ansdell2017}, and thus may vary in wide limits. However, this may not be the only cause of mismatch between the intrinsic and synthetic disk masses. Dust opacity is known to depend on the dust size spectrum and, in particular, on the maximum dust size. Since dust growth has proven to be efficient already in the very early stages of evolution, uncertainties with the actual sizes of dust grains and, thus, with the dust opacity may also affect the dust mass estimates.
Instead of using the OH5 opacity $\kappa_{\lambda=1.3~\mathrm{mm}}=0.89$~cm$^2$~g$^{-1}$  from \citet{OssenkopfHenning1994}, as was done in Sect.~\ref{Sect:disk-mass} following the common practice in observational astronomy \citep{Tobin2020,Kospal2021},
we now adopt the dust-size dependent absorption and scattering opacities from \citet{Woitke2016} (see Fig.~\ref{fig:dustOp} in Appendix~\ref{App:opacity}). 
We note that \citet{OssenkopfHenning1994} opacities are proper to use for cold dark clouds, rather than for very high-density environments of disks.

Figure~\ref{fig:dust-mass} shows the synthetic dust masses in the disk $M_{\rm 90\%}^{\rm obs}$, derived using Equation~(\ref{Eq:mass}) for a wide range of dust temperatures and maximum dust sizes. To guide the eye, the black line shows the intrinsic dust mass in the disk, $M_{\rm disk}^{\rm mod}=673~M_\oplus$. The slope of the dust size distribution is always kept equal $p=-3.5$ and the distance to the source is 140~pc.

Clearly, there is no universal dust temperature that would provide the best match between the synthetic and intrinsic dust masses in the disk if dust growth is considered and the corresponding dust-size-dependent opacities are used. The dust temperature for which both masses match ($M_{\rm 90\%}^{\rm obs}=M_{\rm disk}^{\rm mod}=673~M_\oplus$) for a particular dust size distribution varies from $\approx 23$~K in disks with low to moderate dust growths ($a_{\rm max}\le 100$~$\mu$m) to as low as 3.0~K for disks with $a_{\rm max}\approx 1.0$~mm, and again increases to approx 8--20~K for disks with advanced dust growth ($a_{\rm max} \ge 10$~cm). Extremely low dust temperatures required to match the intrinsic dust mass  around $a_{\rm max}\approx 1.0$~mm are caused by the increase in dust  opacity at the opacity cliff (see Fig.~\ref{fig:dustOp}). We also note that an assumption of $T_{\rm d}>23$~K would always underestimate the intrinsic dust mass, regardless of the progress in dust growth, while lower temperatures may also overestimate dust mass in the limit of $a_{\rm max}\le 100$~$\mu$m.

\begin{figure}  
    \centering
    \includegraphics[width=0.6\linewidth]{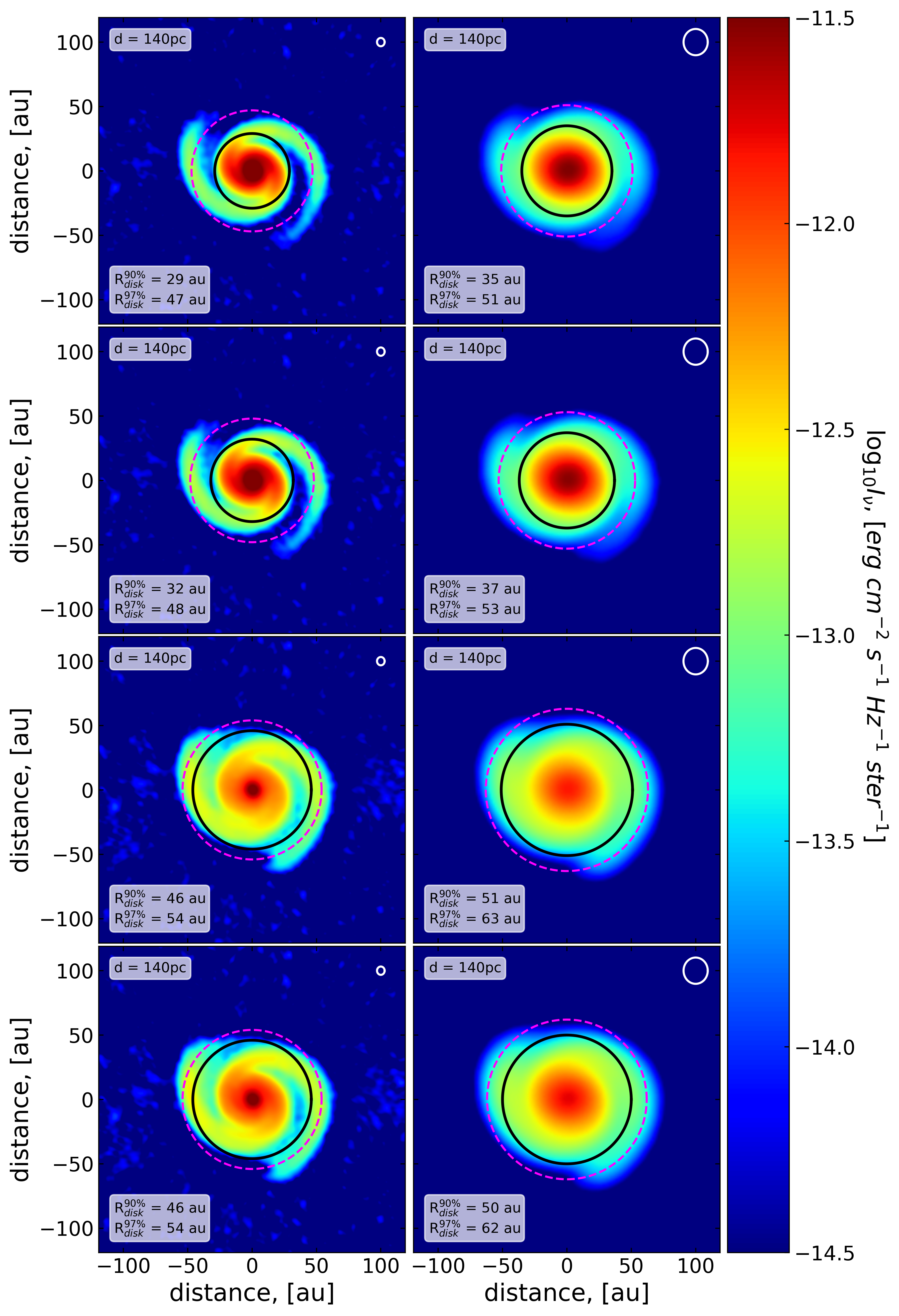}
    \caption{Synthetic intensity distribution of the model disk at 1.3~mm postprocessed using the ALMA OST for Band~6.  The two columns show the effects of different beam sizes: $0.042^{\prime\prime} \times 0.046^{\prime\prime}$ (left) and $0.134^{\prime\prime} \times 0.146^{\prime\prime}$ (right). The distance to the object is 140~pc in both cases. The models shown from top to bottom: $a_{\rm max}^{\rm disk} = 10 \ \mu$m, $100 \ \mu$m, $1$~mm, and variable maximal size in the disk. The black and red circles cover the regions of the disk where 90\%  and 97\% of the total flux is contained. The linear size of the beam is indicated with white circles.}
    \label{fig:resolution}
\end{figure}

\subsection{Effects of ALMA configuration}
In the previous sections, model disks located at different distances of 140~pc, 350~pc, and 700~pc but observed with a similar beam size of $0.042^{\prime\prime} \times 0.046^{\prime\prime}$  were considered. 
This setup imitated distinct star-forming clusters located at different distances from the Sun but observed with the same ALMA configuration.
In this section, we explore another setup and consider a model disk observed at a fixed distance of 140~pc but with different angular resolution. In particular, two beam sizes are considered: $0.042^{\prime\prime} \times 0.046^{\prime\prime}$ and $0.134^{\prime\prime} \times 0.146^{\prime\prime}$.  The linear sizes of the beam in this case are 6~au and 16~au, respectively. This case imitates observations of the same star-forming cluster but with different ALMA configurations (e.g., Cycle~11, Configurations C-8 and C-6, respectively).

\begin{figure}  
\begin{centering}
\includegraphics[width=0.85\linewidth]{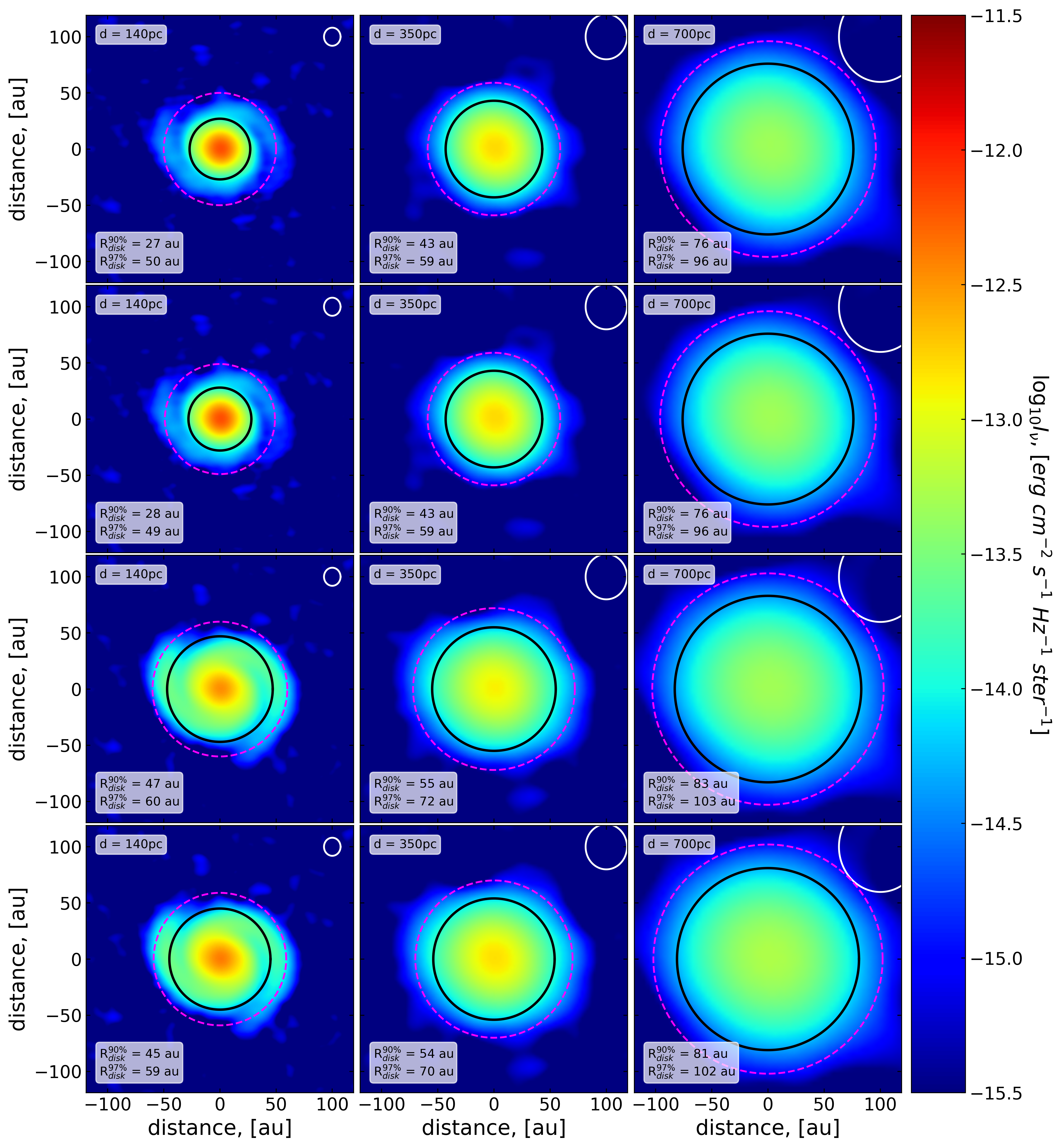}
\par \end{centering}
\caption{Similar to Fig.~\ref{fig:6alma} but for Band 3 with a beam size of $0.105^{\prime\prime} \times 0.115^{\prime\prime}$.} 
\label{fig:3alma}
\end{figure}

Figure~\ref{fig:resolution} compares the resulting synthetic disk images. The effect of changing the beam size on the synthetic disk masses and sizes is similar to that seen in Figure~\ref{fig:6alma} where the beam size was fixed but distance to the source was changing. In particular, the synthetic disk radius grows with increasing beam size by about 10\%, likely due to the effect of beam smearing, while the dust mass is only slightly affected and increases by less than 1\%. It seems that the beam smearing insignificantly alters the total flux, which determines the dust mass estimates (see Eq.~\ref{Eq:mass}). Although the disk radius increases, this effect has little consequence for the mass estimates because the radiation intensity rapidly declines with radial distance near the disk outer edge (see, Fig.~\ref{fig:R-disk2}). 
We also note that the spiral pattern is  significantly smoothed out and barely visible for a beam size of $0.134^{\prime\prime} \times 0.146^{\prime\prime}$, even for the closest possible distance of 140~pc.

\subsection{Other ALMA bands}
In this section, we show synthetic observations of the disk at the ALMA Band 3 ($\lambda=3.0$~mm) for comparison. The beam size in this case is set equal to $0.105^{\prime\prime} \times 0.115^{\prime\prime}$ and the corresponding configuration provides an angular resolution that is a factor $\approx 2.5$ lower than the maximum achievable resolution in  Band~3. The linear sizes of the beam at the adopted distances of 140~pc, 350~pc, and 700~pc are $\approx~$15~au, 38~au, and  75~au.
Dust opacities in \citet{OssenkopfHenning1994} were tabulated only out to $\lambda=1.3$~mm. For consistency, we extrapolate the opacity to longer wavelengths using the following relation 
\begin{equation}
\kappa_{\lambda=3.0~\mathrm{mm}} = \kappa_{\lambda=1.3~\mathrm{mm}} \left({\lambda=1.3~\mathrm{mm} \over \lambda=3.0~\mathrm{mm}}\right)^\beta,
\label{eq:opac}
\end{equation}
where the dust opacity index is set equal to $\beta=1.8$ as in \citet{OssenkopfHenning1994}.

Figure~\ref{fig:3alma} presents the resulting synthetic images. The radiation intensity in Band~3 notably decreases compared to  Band~6, as expected for a longer wavelength. 
At the same time, the integrated flux increases with growing $a_{\rm max}$ as Table~\ref{table:2_b3} demonstrates, unlike Band~6 where we saw the opposite trend caused by flux dilution due to a sharp increase in dust scattering around $a_{\rm max}=1.0$~mm. Apparently, the increase of radiation flux due to transition from the optically thin regime to the optically thick one (see Fig.~\ref{fig:OptDepth} below) outweighs the effect of dust scattering in Band~3.

\begin{figure}
\begin{centering}
\includegraphics[width=\linewidth]{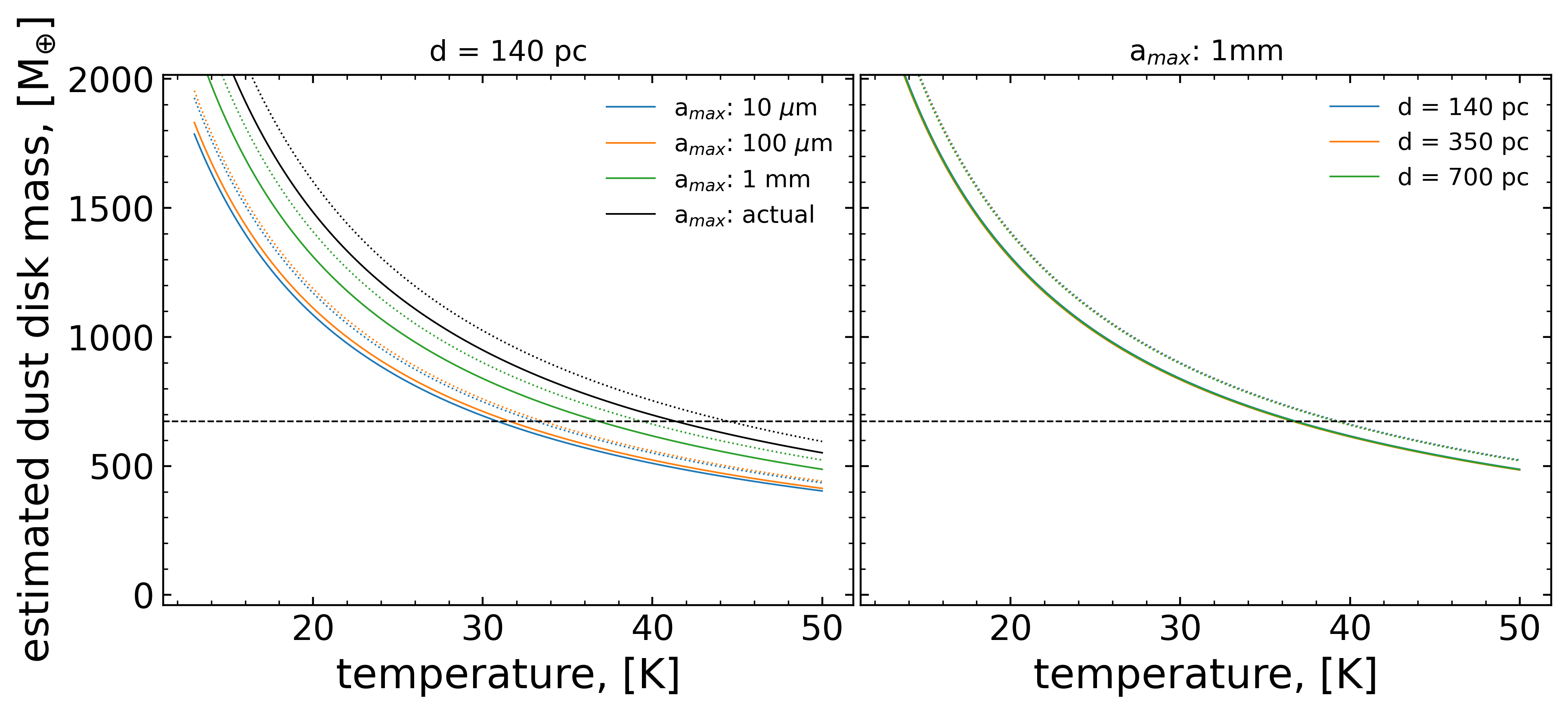} 
\par \end{centering}
\caption{Similar to Fig.~\ref{fig:newmassB6} but for Band~3.}
\label{fig:newmassB3}
\end{figure} 

\begin{table}
\center{\caption{\label{table:2_b3} Integral radiation fluxes in Band 3}}
\begin{tabular}{  c c c c }
\hline \hline
Dust size & $F_{90\%}^{\rm obs} / F_{97\%}^{\rm obs}$ & $F_{90\%}^{\rm obs} / F_{97\%}^{\rm obs}$  & $F_{90\%}^{\rm obs} / F_{97\%}^{\rm obs}$\\
$[a_{\rm max}]$ & [mJy] & [mJy] & [mJy]  \\ [0.5ex]
\hline \\ [-2.0ex]
10~$\mu$m & $37.6 / 40.6$ & $6.02 / 6.47$ & $1.50 / 1.62$ \\
100~$\mu$m & $38.5 / 41.1$ & $6.11 / 6.57$ & $1.53 / 1.64$ \\
1.0~mm & $45.4 / 48.8$ & $7.23 / 7.78$ & $1.81 / 1.94$  \\
actual & $51.4 / 55.5$ & $8.25 / 8.84$ & $2.05 / 2.21$ \\ [1.0ex]
\hline
\end{tabular}
\center{ \textbf{Notes.} Similar to Table~\ref{table:2}, but for Band 3.} 
\end{table}

We find that the synthetic disk radius increases with distance somewhat stronger than in the case of Band~6. This can be expected, as the linear sizes of the beam in Band~3 are greater than for the case of Band~6 (the latter is generally characterized by higher resolution) and the effect of beam smearing in Band~3 is stronger. The increase in  $R_{90\%}^{\rm obs}$  and $R_{97\%}^{\rm obs}$ can be as large as 70\% for $a_{\rm max} \le 100~\mu$m and about 90\% for bigger grains.
We also see an increase in the synthetic disk radius as dust grows. The effect is more pronounced for smaller linear sizes of the beam (better resolution) where the effect of beam smearing is not dominant.  The resulting synthetic disk radii are provided in Table~\ref{table:6}.  We note here that the disk is marginally resolved at 350~pc and is not resolved at 700~pc, thus the corresponding disk size estimates must be taken with care.

The synthetic dust masses are shown in Figure~\ref{fig:newmassB3} for a range of $T_{\rm d}$ and \citet{OssenkopfHenning1994} opacity. A good agreement with the intrinsic dust mass is achieved for $T_{\rm d}=30.9$~K, 31.6~K, 36.8~K and 41.4~K in models with $a_{\rm max}=10$~$\mu$m, 100~$\mu$m, 1.0~mm and in the model with the actual dust size distribution, respectively. All estimates adopt $R^{\rm obs}_{90\%}$ as the disk radius. These best-fit $T_{\rm d}$ are within the range of average dust temperatures adopted in observational studies \citep{Ansdell2017,Tobin2020,Kospal2021}. This means that the dust mass estimates in Band~3 are expected to be more accurate than in Band~6, although notable deviations, both underestimates and overestimates, are possible as $T_{\rm d}$ varies in the 20--50~K limits. In general, the best-fit $T_{\rm d}$ increases at the advanced stages of dust growth. As was the case for Band~6, the synthetic disk masses weakly depend on the distance to the source. 
The resulting synthetic disk masses for several $T_{\rm d}$ are provided in Table~\ref{table:5}.

\begin{figure} 
    \centering
    \includegraphics[width=1.0\linewidth]{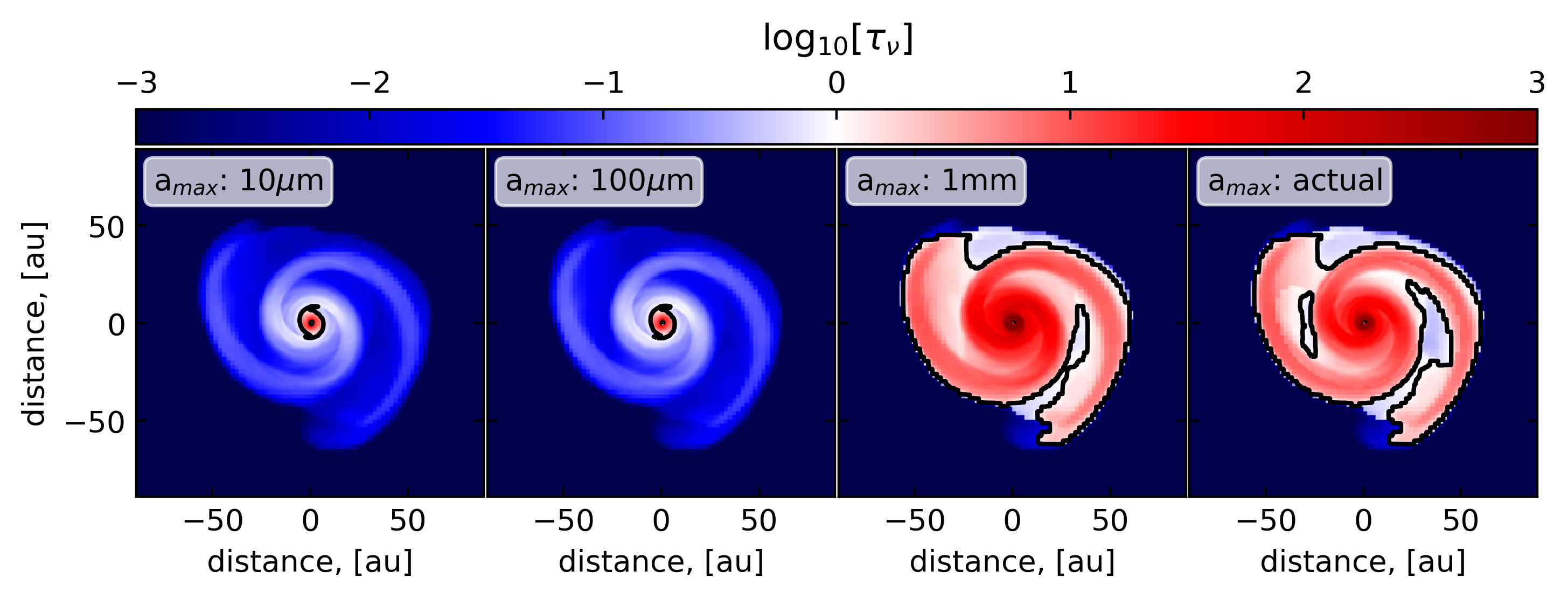}
    \caption{Similar to Fig.~\ref{fig:3} but for Band 3.}
    \label{fig:OptDepth}
\end{figure}

The reason for a better recovery of the intrinsic dust mass is in the systematically lower optical depths of the disk in Band~3. Figure~\ref{fig:OptDepth} shows the optical depth in Band~3 for models with different maximum dust sizes $a_{\rm max}$, considering both dust absorption and scattering. For models at the initial stages of dust growth ($a_{\rm max}\le 100$~$\mu$m), the disk is mostly optically thin, apart from the very inner region with a size of several astronomical units. For advanced stages of dust growth ($a_{\rm max}\ge 1.0$~mm), the bulk of the disk is optically thick, but on average $\tau_{\rm \nu}$ is a factor of several lower than for the Band~6 case (see Fig.~\ref{fig:3}).

\begin{table}
\center{\caption{\label{table:5}Synthetic dust masses in Band~3 for particular $T_{\rm d}$ }}
\begin{tabular}{  c c c c }
\hline \hline
Dust size & $M_{90\%}^{\rm obs}(T_{\rm d}=20$~K) & $M_{90\%}^{\rm obs}(T_{\rm d}=30$~K)  & $M_{90\%}^{\rm obs}(T_{\rm d}=40$~K)\\
$[a_{\rm max}]$ & [$M_\oplus$] & [$M_\oplus$] & [$M_\oplus$]  \\ [0.5ex]
\hline \\ [-2.0ex]
10~$\mu$m & 1085 & 694 & 510 \\
100~$\mu$m & 1112 & 711 & 522 \\
1.0~mm & 1311 & 839 & 616  \\
actual & 1483 & 949 & 697 \\ [1.0ex]
\hline
\end{tabular}
\center{ \textbf{Notes.} Distance to the source is $d=140$~pc.  The intrinsic dust mass in the disk is $M_{\rm dust}^{\rm mod}=673~M_\oplus$. }
\end{table}

\begin{figure}  
\begin{centering}
\includegraphics[width=0.8\linewidth]{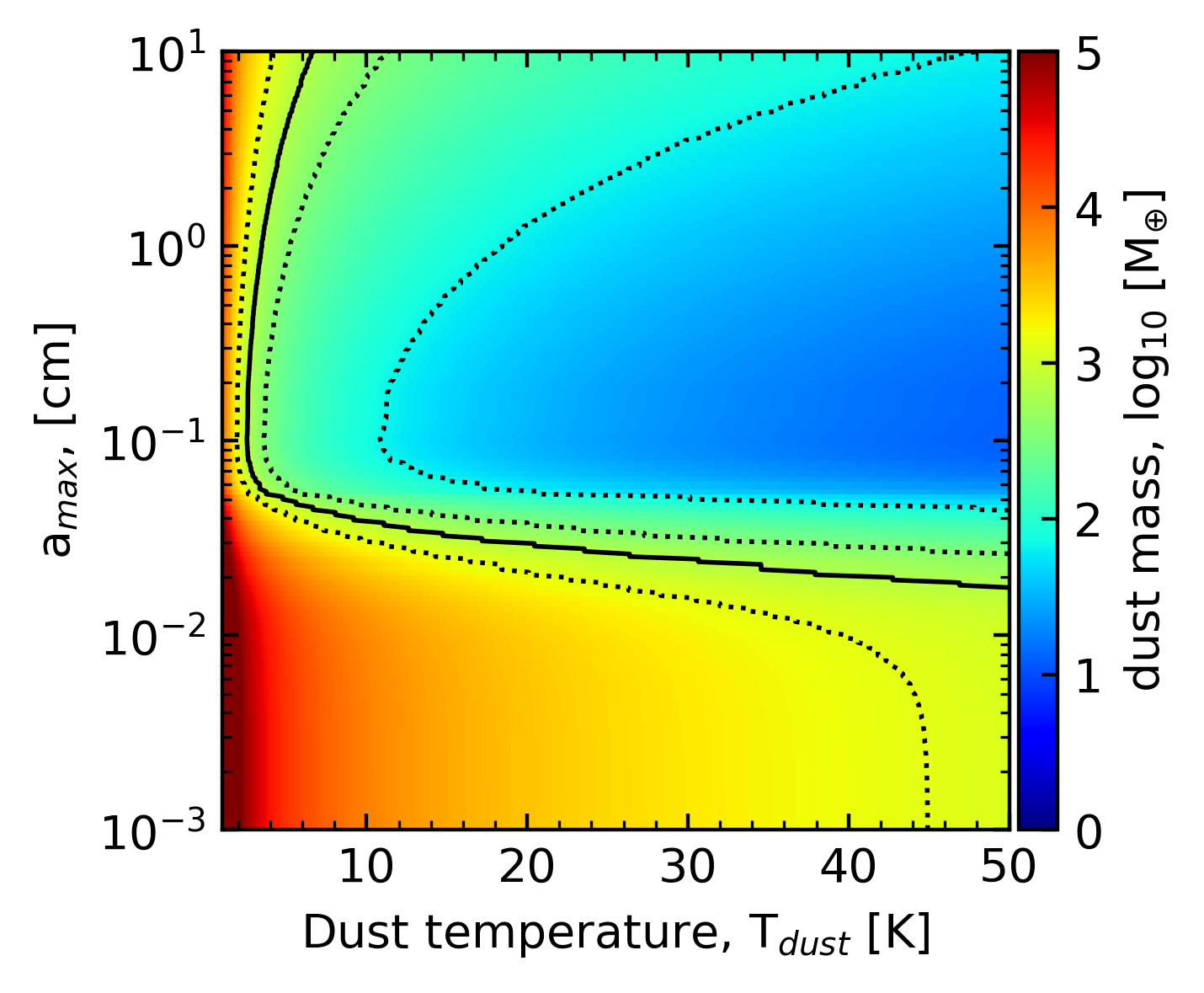}
\par \end{centering}
\caption{Similar to Fig.~\ref{fig:dust-mass-B3} but for Band~3.
} 
\label{fig:dust-mass-B3}
\end{figure}

Figure~\ref{fig:dust-mass-B3} shows the effects of dust growth and varying dust temperature on dust mass estimates. Now, we use the dust-size-dependent opacities (rather than those of \citealt{OssenkopfHenning1994}) to calculate the dust mass using~Equation~(\ref{Eq:mass}). As for the case of Band~6, there is no universal dust temperature that can provide a good match between synthetic and intrinsic dust masses in the disk when dust growth is considered. For the initial stages of dust growth, when $a_{\rm max} \le \mathrm{a~few} \times 100$~$\mu$m, any choice of $T_{\rm d}$ below 50~K overestimates the true dust mass. 
At larger $a_{\rm max}$, the dust mass is in general underestimated, unless for very small $T_{\rm dust}< 10$~K.

We also note that the spiral structure is only detectable at a distance of 140~pc and is completely smoothed out at larger distances, unlike Band~6 for which the spiral structure was still visible at a distance of 350~pc (see Fig.~\ref{fig:6alma}). The $R_{90\%}^{\rm obs}$ circle cuts out a large fraction of the spiral arms for models with $a_{\rm max}\le 100$~$\mu$m, as was also the case for Band~6.

\begin{table*}
\center{\caption{\label{table:6}Synthetic disk radii}}
\begin{tabular}{  c | c c | c c | c c}
\hline \hline
Dust size & $R_{90\%}^{\rm obs} / R_{97\%}^{\rm obs}$ & 
$\triangle R_{90\%}^{\rm obs} / \triangle R_{97\%}^{\rm obs}$ &
$R_{90\%}^{\rm obs} / R_{97\%}^{\rm obs}$  &  $\triangle R_{90\%}^{\rm obs} / \triangle R_{97\%}^{\rm obs}$  & $R_{90\%}^{\rm obs} / R_{97\%}^{\rm obs}$ & $\triangle R_{90\%}^{\rm obs} / \triangle R_{97\%}^{\rm obs}$  \\
$[a_{\rm max}]$ & [au] & [au] & [au] & [au]  & [au] & [au] \\ [0.5ex]
\hline \\ [-2.0ex]
10~$\mu$m & $27/50$ & -26/-3 & $43/59$ & -10/+6 & $76/96$ & +23/+43 \\
100~$\mu$m & $28/49$ & -25/-4 & $43/59$ & -10/+6 & $76/96$ & +23/+43 \\
1.0~mm & $47/60$ & -6/+7 & $55/72$ & +2/+19 & $83/103$ & +30/+50  \\
actual & $45/50$ & -8/+7 & $54/70$ & +1/+17 & $81/102$ & +28/+49 \\ [1.0ex]
\hline
\end{tabular}
\center{ \textbf{Notes.} Similar to Table~\ref{table:1} but for Band~3} 
\end{table*}

\section{Discussion}
\label{Sect:discuss}

This work is focused on the earliest stages of disk evolution, when the FHSC or the protostar has not yet provided sufficient luminosity output to start dominating the energy balance in the disk and the disk temperature is instead set by internal hydrodynamic processes. This makes the radiative transfer simulations somewhat easier as there is no uncertainty with the intrinsic parameters of the growing star (such as the effective temperature and stellar radius) and the stellar luminosity output such as accretion luminosity. At the same time, this stage is short and is observationally difficult to capture. There are very few FHSC candidates \citep{2010ApJ...715.1344C,Pezzuto2012,Duan2023}, and none have been reliably verified to date. Nevertheless, we think that it is important to explore these earliest stages, even though this study may currently represent only a theoretical prospective. In a follow-up study, we will explore subsequent Class 0 and I stages of disk evolution,  where the protostar significantly contributes to the disk energy balance, with a similar numerical setup of ngFEOSAD. The current proof-of-concept theoretical work provides an important bridge to more observationally motivated studies of the Class 0 and I stages, including the disk inclination effects which were omitted in the present work.

We found that that similar disks but located in distinct star-forming regions at different distances from the Sun can have different apparent disk radii and the variation around the intrinsic value may be substantial, up to a factor of two, depending on the dust growth stage in the disk. When using the \citet{OssenkopfHenning1994} opacities and dust temperatures typical for Class~0 and I disks, Equation~(\ref{Eq:mass}) can yield significant dust mass underestimates by factors of several. The problem is more severe in Band~6 than in Band~3 of ALMA. In the latter case, the dust mass can actually be even overestimated. We also demonstrated that by adjusting the rather uncertain dust temperature in Equation~(\ref{Eq:mass}), it becomes possible to recover the intrinsic mass of dust in the disk. The best-fit values of $T_{\rm d}$ strongly depend on the ALMA band, but are weakly sensitive to the distance to the source.

The situation becomes more complicated when the dust-size-dependent opacities are considered. The best-fit $T_{\rm d}$ begin to strongly depend on $a_{\rm max}$, that is, on the dust growth stage in the disk. A good match between the synthetic and intrinsic dust masses for $a_{\rm max} \ge 1.0$~mm may require a dust temperature that is much lower than usually adopted in observational studies. We note that the best-fit $T_{\rm d}$ also depends on $a_{\rm max}$ via the corresponding variations in the integrated radiation flux (see Tables~\ref{table:2} and \ref{table:2_b3}), but the dependence via the varying dust opacity is stronger.
We also note that the intricate relation between $T_{\rm d}$ and $a_{\rm max}$ implies some dependence of the former on the latter. This is possible considering that $T_{\rm d}$ depends on dust opacity and dust opacity in turn depends on $a_{\rm max}$. In addition, the gas-to-dust energy transfer rate depends on the dust properties, such as the total surface area which changes as dust grows, but these effects are beyond the present study.

Our model and synthetic disk radii (see Table~\ref{table:1}) are broadly consistent with those inferred for Class~0 disks by the CAMPOS and VANDAM surveys of the Aquila, Corona Australis, Ophiuchus North, Ophiuchus, Orion, and Serpens molecular clouds \citep{Tobin2020,Hsieh2024}. Our model, however, considers the disk at the FHSC to protostar transition and may not be representative in terms of disk mass and radius due to short duration of this phase.
In any case, the comparison of the synthetic disk radii and masses with the observationally derived samples was not the purpose of this study. Here, we explored the accuracy of inferring the true disk radii and masses from synthetic millimeter fluxes, excluding the associated uncertainties with the luminosity of the growing star but taking dust growth into consideration.

The mismatch between intrinsic and synthetic disk masses was also reported by \citet{2014DunhamVorobyov} in the context of Class 0 and I disks, though without considering the effects of dust growth. It was also found that the disk masses may be underestimated by up to factors of 2--3 at millimeter wavelengths and up to an order of magnitude at sub-mm wavelengths, in particular, due to uncertainties in the optical depth and dust temperatures, as was also found in our study. In contrast, \citet{Harsono2015} found that disk masses inferred from synthetic observations in millimeter wavelength agree rather well with the actual disk masses of young embedded disks obtained from magnetohydrodynamics simulations. However, the adopted beam size of 5--15$^{\prime\prime}$ may have encompassed the large-scale emission of the surrounding envelope, rather than the disk itself.  They also retrieved the disk mass from the total envelope+disk flux (eqs.~1-3 in \citealt{2009A&A...507..861J}).  Mass underestimates of gravitationally unstable disks by factors 2.5-30 using synthetic (sub)-mm observations at different ALMA bands was also reported in \citet{Evans2017}.  In a more recent study by \citet{Liu-Flock2022}, focused rather on Class II analytic disk models but with sophisticated dust composition and considering dust growth, possible severe underestimation of dust mass was also reported. The requirement of multi-wavelength observations for more accurate measurements of usually underestimated dust masses was also reported in \citet{Veronesi2025}.
We also note that multi-wavelengths observations of polarized dust emission can help to distinguish between small grains with $a_{\rm max} \le 100$~$\mu$m and larger grains with $a_{\rm max} \sim 1.0$~mm \citep[e.g.,][]{Guillet2020}, thanks to the change in the direction of polarization from perpendicular to parallel with respect to magnetic field lines. Interestingly, this is about a dust size range where a qualitative transition in the synthetic disk images occurs in our model (see Figs.~\ref{fig:3} and \ref{fig:6alma}). 
To conclude, most theoretical studies report a substantial underestimate of the dust mass when using the conventional techniques (see Eq.~\ref{Eq:mass}), and this issue may be inherent to all stages of disk evolution, including the very earliest considered in this work.

We finally note that the magnetic field and turbulence can have significant impacts on disk masses and sizes, especially in disk's early stages of evolution \citep{Seifried2012,Santos-Lima2013,Lam2019,Maury2022}.  For instance, magnetic braking and magnetic disk winds can reduce the disk mass and size, and magnetohydrodynamic instabilities can shape the infalling envelope into streamers instead of axially symmetric infall \citep{MachidaBasu2025}.  While the peculiar structure of the infalling envelope is unlikely to affect the synthetic images at mm-wavelengths because of its rarefied nature, a reduced disk mass could also result in lower optical depths, and, as a consequence, in more accurate mass estimates via Eq.~(\ref{Eq:mass}). This effect can be mimicked by our collapse simulations with lower initial cloud core masses, which we leave for future studies.

\section{Conclusions}
\label{Sect:conclude}
In this work, we investigated the accuracy with which the mass and radius of a young protoplanetary disk can be inferred using its dust thermal emission at millimeter wavelengths. We started with a three-dimensional hydrodynamic simulation of protoplanetary disk formation with the ngFEOSAD code. We then exported the resulting density and temperature distributions of dust in the disk and envelope into the RADMC-3D code, using either the realistic dust size distribution derived directly from hydrodynamic simulations, or making simplifying assumptions about the maximum dust size to explore the possible effects of dust growth. Next, we postprocessed the radiation fluxes with the ALMA observational support tool to generate realistic synthetic images of our model disk at different ALMA bands and with different resolution.  
These images were finally used to calculate the synthetic disk radii and the dust masses contained with these radii using the conventional method (see Eq.~\ref{Eq:mass}) adopted in observational astronomy.

Four models with the maximum dust size corresponding to moderate and advanced stages of dust growth were considered:  $a_{\rm max}=10$~$\mu$m, $a_{\rm max}=100$~$\mu$m, $a_{\rm max}= 1.0$~mm, and the model with the spatially varying distribution of $a_{\rm max}$ obtained from hydrodynamic simulations. We focused on a very young protoplanetary disk 
that formed in our simulations around a first hydrostatic core, just before it transits to a protostar. This allowed us to exclude uncertainties with the protostellar radiation output and focus on other parameters that may influence the disk mass and radius measurements. 
The main findings can be summarized as follows.
\begin{itemize}

    \item When choosing the \citet{OssenkopfHenning1994} opacity, dust mass can be underestimated by factors of $1.4-4.2$ in Band~6 for the mean dust temperatures $T_{\rm d}=20-40$~K, a typical range adopted in observational studies.
A good match between the synthetic and intrinsic dust masses in Band~6 can be recovered for dust temperatures $T_{\rm dust}=13-16$~K. Observations in Band~3 can both underestimate or overestimate the dust mass by up to a factor of two depending on the choice of $T_{\rm d}$. The best-bit $T_{\rm d}$ lie in the 31-41~K limits.
Synthetic disk masses in ALMA Bands~3 and 6 seem to be weakly sensitive to the distance to the source.

\item When more realistic dust-size-dependent opacities are considered (see Fig.~\ref{fig:dustOp}), the discrepancy between the synthetic and intrinsic dust masses begins to strongly depend on the maximum dust size, in addition to strong dependence on $T_{\rm d}$. To achieve a fair agreement, information on the maximum dust size in the disk is desirable, and  extremely low values of dust temperature may be required to reconcile the synthetic and intrinsic masses for $a_{\rm max} \ge 1.0$~mm.

    \item Synthetic disks look bigger at the advanced stages of dust growth $a_{\rm max}\ge 1.0$~mm. This may be due to the fact that most of the disk becomes optically thick as dust grows and the dust opacity reaches a peak value near $\lambda= a_{\rm max}/ (2 \pi)$.

    \item Synthetic disk sizes grow with increasing distance to the source and deteriorating linear resolution, likely due to the effect of beam smearing.

    \item Dust scattering becomes significant at the advanced stages of dust growth ($a_{\rm max} \simeq 1.0$~mm), affecting the radial intensity profiles and dust mass estimates, in particular in Band~6. 

    \item  Spiral pattern generated by gravitational instability is easier detected at the early stages of dust growth ($a_{\rm max}\le 100~\mu$m) than at the advanced stages ($a_{\rm max} \ge 1.0$~mm) and in Band~6 rather than in Band~3.

\end{itemize}

Both adopted definitions for the disk radius,  90\% or 97\% of the total flux, can either underestimate or overestimate the intrinsic disk radius, depending on the linear resolution of observations and maximum dust size. This means that a young protoplanetary disk viewed at different resolution and at different stages of dust growth would have different apparent disk radii, and the variation around the intrinsic value can be substantial, up to a factor of two.
The effect of beam smearing with increasing distance can be offset by subtracting about 1/2 of the linear size of the beam from the synthetic disk radii.
This work provides an important bridge to more observationally accessible Class 0 and I stages of disk evolution. In a follow-up study, we plan to use the developed algorithm to study the accuracy with which Class 0 and I disk masses and sizes can be recovered using thermal dust emission.

\section*{Conflict of Interest Statement}

The authors declare that the research was conducted in the absence of any commercial or financial relationships that could be construed as a potential conflict of interest.

\section*{Author Contributions}
E.I.V suggested the main contents of the paper, performed ngFEOSAD simulations and wrote the paper. A.M.S. performed RADMC-3D simulations and prepared most figures. V.G.E. calculated the disk masses and radii from hydrodynamic simulations. M.D. and M.G. provided valuable inputs to the interpretations of synthetic disk models.

\section*{Funding}
This work was supported by the Austrian Science Fund (FWF), project I4311-N27, DOI:10.55776/I4311.
A.M.S. and V.G.E. also acknowledge support by the Ministry of Science and Higher Education of the Russian Federation (State assignment in the field of scientific activity 2023, GZ0110/23-10-IF).

\section*{Acknowledgments}
We are thankful to the anonymous referees for constructive comments that helped to improve the manuscript.
Simulations were performed on the Austrian Scientific Cluster (\href{https://vsc.ac.at/}{https://asc.ac.at/})

\section*{Data Availability Statement}
All data related to this study can be provided on reasonable request.

\bibliographystyle{Frontiers-Harvard} 
\bibliography{refs}

\newpage

\section{Appendix A. Basic model equations}
\label{App:equations}

The dynamics of gas is followed by solving the equations of continuity and momentum
\begin{equation}
\label{eq:cont}
\frac{{\partial \rho_{\rm g} }}{{\partial t}}   + \nabla  \cdot 
\left( \rho_{\rm g} {\bl v} \right)  = 0,  
\end{equation}
\begin{equation}
\label{eq:mom}
\frac{\partial}{\partial t} \left( \rho_{\rm g} {\bl v} \right) +  \nabla \cdot \left( \rho_{\rm
g} {\bl v} \otimes {\bl v} + \mathbb{I} P \right)   =    - \rho_{\rm g} \, \nabla \Phi 
 - \rho_{\rm d,gr} \, {\bl f},
\end{equation}
where $\rho_{\rm g}$ is the volume density of gas, $\Phi$ is the gravitational potential, $\bl v$ is the gas velocity,  $P$ is the gas pressure, $\mathbb{I}$ is the unit tensor, and $\bl f$ is the friction force per unit dust mass between gas and dust. We used a barotropic equation of state for the closure relation between the gas pressure and density, which smoothly links an isothermal envelope with the disk and accounts for rotational degrees of freedom of H$_2$ at temperatures exceeding $\approx 100$~K \citep[see][for details]{VorobyovKulikov2024}. The full energy balance equation is therefore not solved.

Dust is considered as a pressureless fluid  having two populations: small dust that is dynamically linked to gas and grown dust that is dynamically decoupled from gas. The corresponding dust dynamics equations are
\begin{equation}
\label{eq:cont_dust_small}
\frac{{\partial \rho_{\rm d,sm} }}{{\partial t}}   + \nabla  \cdot 
\left( \rho_{\rm d,sm} {\bl v} \right)  = - S(a_{\rm max}),  
\end{equation}
\begin{equation}
\label{eq:cont_dust_grown}
\frac{{\partial \rho_{\rm d,gr} }}{{\partial t}}   + \nabla  \cdot 
\left( \rho_{\rm d,gr} \, {\bl u} \right)  = S(a_{\rm max}),  
\end{equation}
\begin{eqnarray}
\label{eq:mom-dust}
\frac{\partial}{\partial t} \left( \rho_{\rm d,gr}\, {\bl u} \right) +  \nabla \cdot \left( \rho_{\rm
d,gr} \, {\bl u} \otimes {\bl u}  \right)  &=&     - \rho_{\rm d,gr} \, \nabla \Phi + \rho_{\rm d,gr} \, {\bl f}  \nonumber \\ 
&+&  S(a_{\rm max}) \, {\bl v},
\end{eqnarray}
where $\rho_{\rm d,sm}$ and $\rho_{\rm d,gr}$ are the volume densities of small and grown dust, respectively, $\bl u$ the velocity of grown dust, and $S(a_{\rm max})$ the conversion rate between small and grown dust populations as dust grows and transits from the small to grown population. We adopt a simple monodisperse model of dust growth according to \citet{Birnstiel2012} and the expression for $S(a_{\rm max})$ can be found in \citet{VorobyovKulikov2024}. 
The dust growth is halted at the fragmentation barrier $a_{\rm frag} = \rho_{\rm g}u^2_{\rm frag}/ (3 \alpha \rho_{\rm s}  v_{\rm th} \Omega_{\rm K})$,
where $u_{\rm frag}=5.0$~m~s$^{-1}$ is a threshold relative velocity of colliding dust grains, above which grain-to-grain collisions lead to destruction rather than to growth, $\Omega_{\rm K}$ is the Keplerian angular velocity, $\alpha=10^{-3}$ is the \citet{1973ShakuraSunyaev} turbulent parameter, $v_{\rm th}$ is the mean thermal velocity of gas, and $\rho_{\rm s}= 3.0$~g~cm$^{-3}$ is the material density of dust grains. 
The drag force between dust and gas is calculated as
\begin{equation}
    {\bl f} = \frac{{\bl v} - {\bl u}}{t_{\rm stop}},
    \label{eq:fric}
\end{equation}
where   $t_{\rm stop}= a_{\rm max} \, \rho_{\rm s} / (\rho_{\rm g} v_{\rm th})$ is the stopping time.

\section{Appendix B. Toomre $Q$-parameter}
\label{App:Toomre}
To analyze the propensity of our model disk to gravitational instability, we calculate the Toomre $Q$-parameter \citep{Toomre1964}
\begin{equation}
    Q ={ {c_{\rm s} \Omega} \over \pi G \Sigma_{\rm g}},
    \label{Eq:Q-par}
\end{equation}
where $c_{\rm s}$ is the sound speed, $\Omega$ the angular velocity, $G$ the gravitational constant, and $\Sigma_{\rm g}$ the surface density of gas. We neglect here a contribution of dust to the sound speed and local surface density, as the dust concentrations above the initial dust-to-gas ratio of 0.01 is small at these early stages of disk evolution. If $Q<1$, the disk becomes unstable and develops a spiral pattern. If $Q<0.6$ in the spiral arms, they can fragment to form compact gravitationally supported clumps \citep{Takahashi2016}.

\begin{figure} 
\begin{centering}
\includegraphics[width=0.8\linewidth]{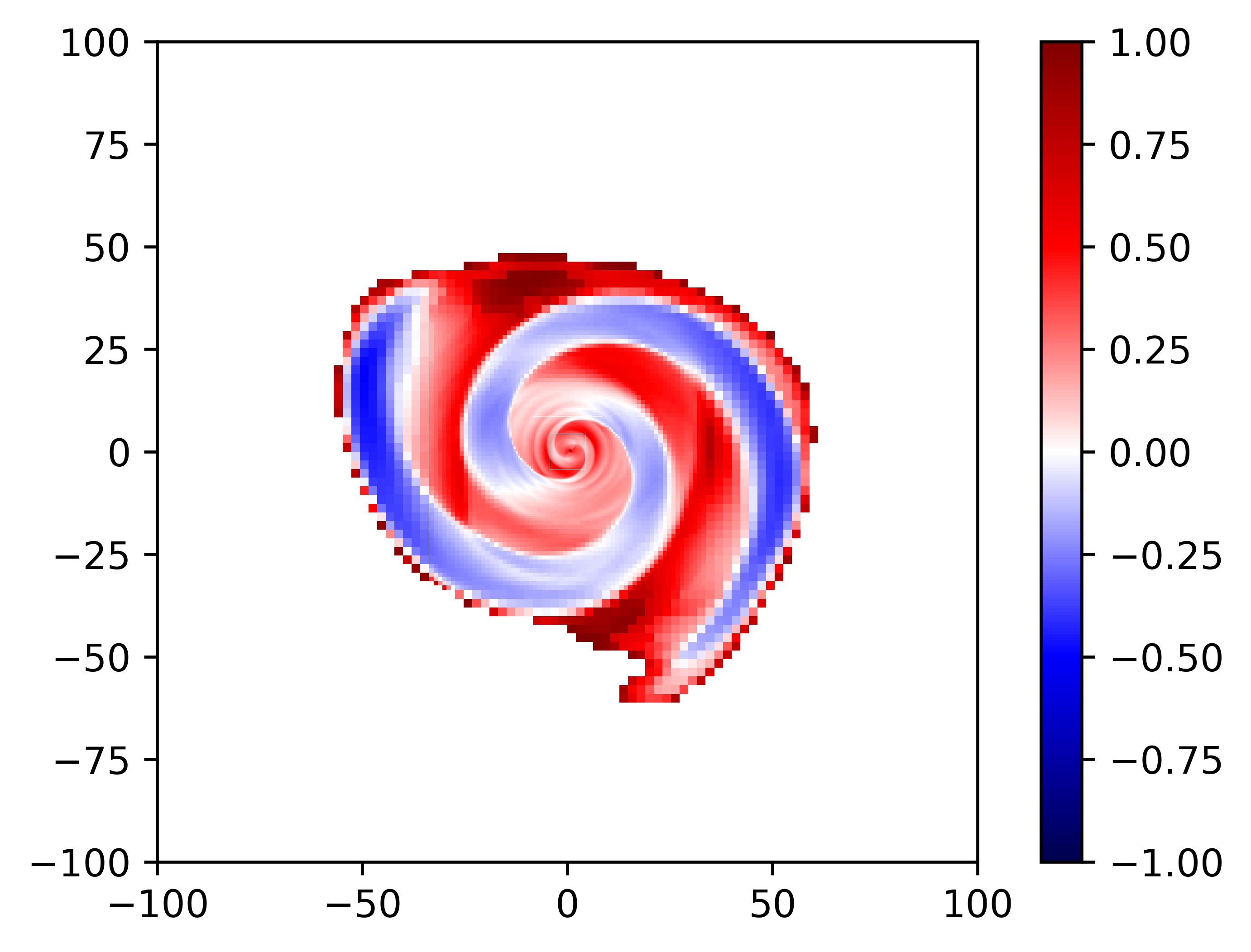} 
\par \end{centering}
\caption{Two-dimensional map of the $Q$-parameter in the disk. The scale bar is in the log units.} 
\label{fig:Q-par}
\end{figure} 

As Figure~\ref{fig:Q-par} demonstrates, the Toomre-parameter in the spiral arms drops below unity, verifying the presence of gravitational instability. However, the $Q$-value also drops below the threshold for gravitational fragmentation in the outer parts of spiral arms ($Q\approx 0.5$), but fragments do not occur. This may be explained by the fact that the threshold criteria were derived for two-dimensional thin disks and require modifications to account for the three-dimensional disk structures. For instance, when deriving $\Sigma_{\rm g}$ we integrate over the entire disk extent rather than over one disk scale height. The use of angular velocity $\Omega$ in Eq.~(\ref{Eq:Q-par}) is also valid only if it is well approximated by the Keplerian velocity, but this may not be so at this early stage of disk evolution in absence of a point-like gravity source.  

\section{Appendix C. Identifying the radius of the model disk}
\label{App:R-disk}
\begin{figure} 
\begin{centering}
\includegraphics[width=0.8\linewidth]{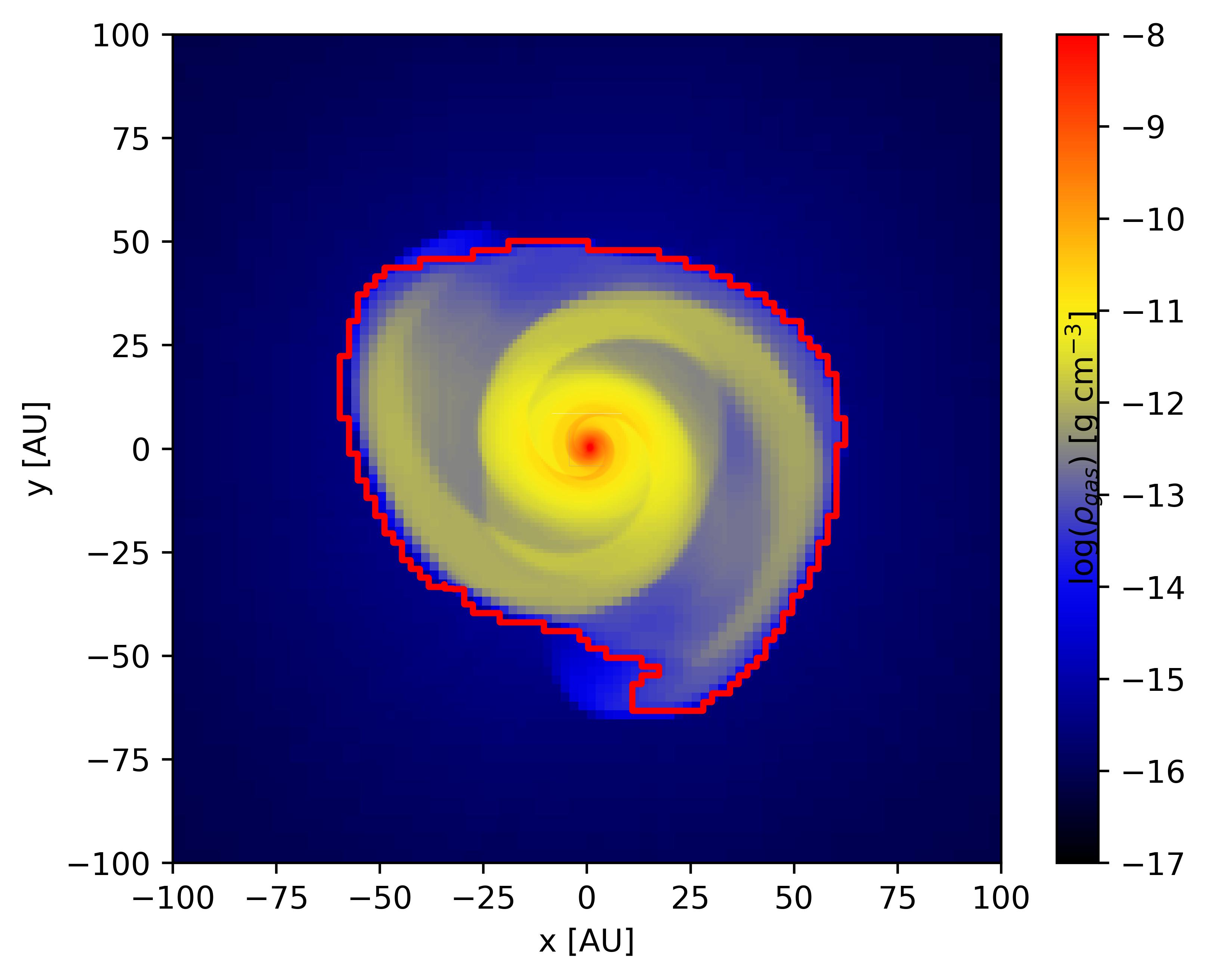} 
\par \end{centering}
\caption{Distinguishing the disk from the infalling envelope. The red contour line indicates the disk outer boundary.} 
\label{fig:disk-R}
\end{figure}

Figure~\ref{fig:disk-R} presents the spatial map of the gas volume density in the disk midplane. The red contour line delineates the disk outer boundary according to the disk tracking mechanism of Sect.~\ref{Sect:disk-track}. As can be seen, the disk tracking mechanism distinguishes the disk from the infalling envelope (black low-density regions) fairly well. The azimuthally averaged disk radius is $R^{\rm mod}_{\rm mod}=53$~au and the maximum extent of the disk at a specific azimuth is 66.5~au.

\section{Appendix D. Effects of dust scattering}
\label{Sect:scatter}

\begin{figure}
\begin{centering}
\includegraphics[width=\linewidth]{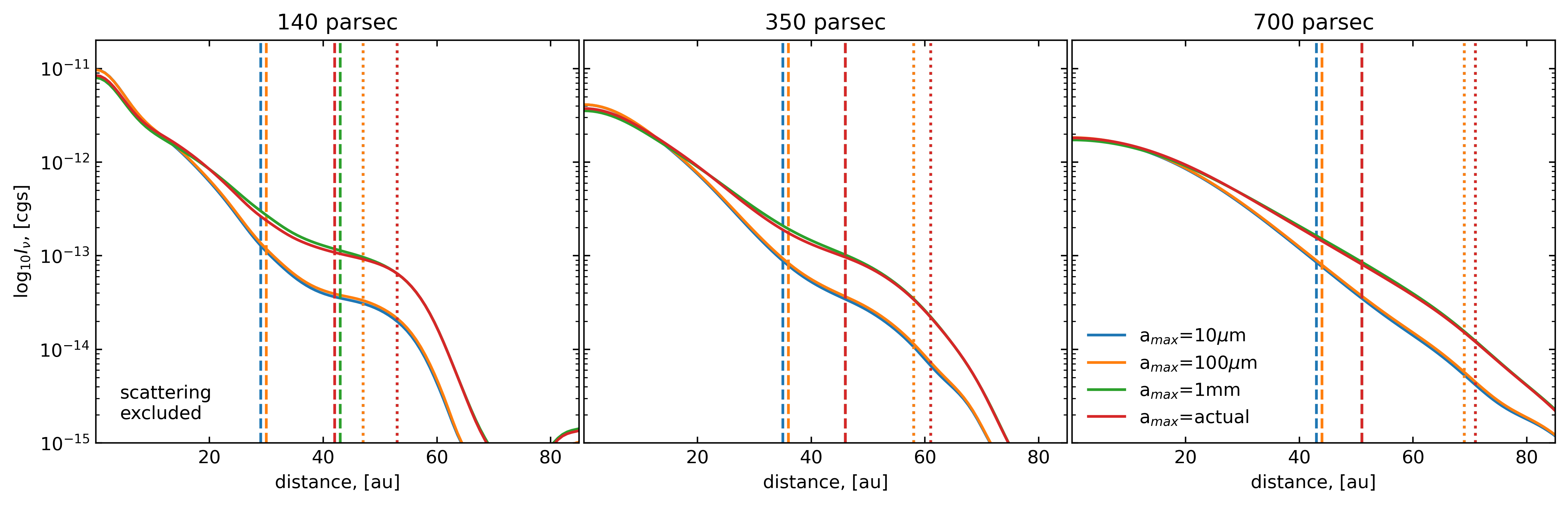} 
\par \end{centering}
\caption{Similar to Fig.~\ref{fig:R-disk2} but only for absorption  dust opacity.} 
\label{fig:R-disk3}
\end{figure}

Figure~\ref{fig:R-disk3} presents the azimuthally integrated radiation intensities in Band~6 obtained for the case of absorption opacity, neglecting the contribution from dust scattering. A comparison with Fig.~\ref{fig:R-disk2} reveals that the radiation intensities of models with advanced dust growth ($a_{\rm max}\ge 1.0$~mm) are characterized by shallower slopes, as in the case when scattering is also considered. However, $I_{\nu}$ in the inner 10-15~au is now of similar magnitude for all models regardless of $a_{\rm max}$, whereas in Fig.~\ref{fig:R-disk2} a significant drop in $I_{\nu}$ for models with $a_{\rm max}\ge 1.0$~mm is evident.  As was shown in \citet{Zhu2019-scatter}, dust scattering modifies the radiation intensity in the optically thick limit as $I_\nu \sim \sqrt{1-\omega_{\nu}} B_{\nu}(T)$, 
where $\omega_{\nu}= \kappa_{\nu, \rm sc}/(\kappa_{\nu, \rm sc}+\kappa_{\nu,\rm abs})$ is the scattering albedo, and  $\kappa_{\nu, \rm sc}$ and $\kappa_{\nu, \rm abs}$ are the scattering and absorption opacities. The former is much smaller than the latter for $a_{\rm max}\le 100$~$\mu$m, but becomes comparable to or grater than the absorption opacity as dust grows to $a_{\rm max}=1.0-10$~mm. This increases the scattering albedo and systematically lowers the radiation intensity and the integrated radiation flux in the advanced stages of dust growth, see Table~\ref{table:2}.

\section{Appendix E. Dust-size-dependent opacities}
\label{App:opacity}
\begin{figure} 
\begin{centering}
\includegraphics[width=0.8\linewidth]{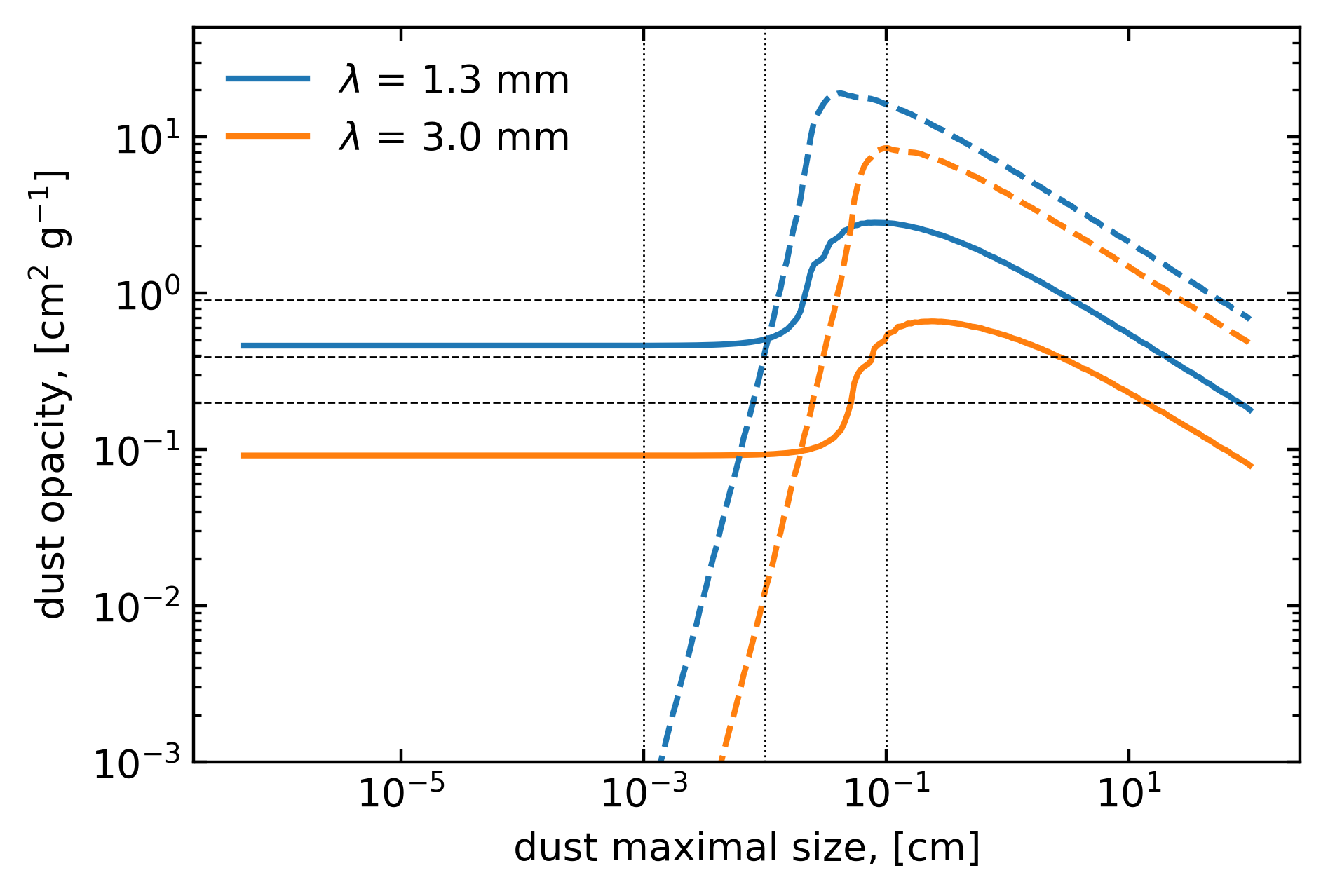} 
\par \end{centering}
\caption{Dust absorption (solid lines) and scattering (dashed lines) opacity as a function of the maximum dust size $a_{\rm max}$. The vertical dotted lines identify the values of $a_{\rm max}$ used in our work. The horizontal dashed lines are the \citet{OssenkopfHenning1994} opacities considered in this work for Band 6 (top line) and Band~3 (middle and bottom lines).} 
\label{fig:dustOp}
\end{figure}

Figure~\ref{fig:dustOp} presents the dust absorption and scattering opacities at $\lambda=1.3$~mm and 3.0~mm calculated with the opTool \citep{Woitke2016}.
In particular, we fix the smallest size of dust grains at $5\times10^{-3}$~$\mu$m, but vary the maximum dust size in the $10^{-3} - 10$~cm limits. The following dust composition is assumed: 40\% -- astrosilicates, 10\% -- foilite, and 50\% -- refractory organics.  A sharp increase at the opacity cliff, $a_{\rm max}=\lambda / (2 \pi)$, is evident.

\end{document}